\documentclass[usenatbib,12pt]{iopart}

\usepackage{iopams}  
\usepackage{graphicx}  
\usepackage{graphics}  
\usepackage{psfrag}
  
\newcommand{\be}{\begin{equation}}      
\newcommand{\ee}{\end{equation}}      
      
\newcommand{\bef}{\begin{figure}}      
\newcommand{\eef}{\end{figure}}      
\newcommand{\bea}{\begin{eqnarray}}    
\newcommand{\eea}{\end{eqnarray}}

\def\lsim{\raise 0.4ex\hbox{$<$}\kern -0.8em\lower 0.62ex\hbox{$\sim$}} 
\def\gsim{\raise 0.4ex\hbox{$>$}\kern -0.7em\lower 0.62ex\hbox{$\sim$}}

\def\f0N{f_0^{(N)}}
\def\bec{\begin{center}}
\def\eec{\end{center}}

\begin{document}

\title[Relaxation to thermal equilibrium in  the self-gravitating sheet model
]
{Relaxation to thermal equilibrium in  the self-gravitating sheet model}

\author[M. Joyce and  T. Worrakitpoonpon]
{M. Joyce${^{1,}}{^{2}}$ and T. Worrakitpoonpon${^{1}}$\\}

\address{
$^1$Laboratoire de Physique Nucl\'eaire et Hautes \'Energies,\\ 
Universit\'e Pierre et Marie Curie - Paris 6,
CNRS IN2P3 UMR 7585,  4 Place Jussieu, 75752 Paris Cedex 05, France\\
$^2$Laboratoire de Physique Th\'eorique de la Mati\`ere Condens\'ee,\\
Universit\'e Pierre et Marie Curie - Paris 6,
CNRS UMR 7600, 4 Place Jussieu, 75752 Paris Cedex 05, France}

\begin{abstract}
We revisit the issue of relaxation to thermal equilibrium in the so-called
``sheet model'', i.e., particles in one dimension interacting by attractive
forces independent of their separation. We show that this relaxation may 
be very clearly detected and characterized by following the evolution of order 
parameters defined by appropriately normalized moments of the phase space distribution
which probe its {\it entanglement} in space and velocity coordinates.
For a class of quasi-stationary states which result from the violent relaxation 
of rectangular waterbag initial conditions, characterized by their virial ratio $R_0$,
we show that relaxation occurs on a time scale which (i) scales approximately  
linearly in the particle number $N$, and  (ii) shows also a strong dependence 
on $R_0$, with quasi-stationary states from colder initial conditions relaxing 
much more rapidly. The temporal evolution of the order parameter
may be well described by a stretched exponential function. We study
finally  the correlation of the relaxation times with the amplitude of fluctuations 
in the relaxing quasi-stationary states, as well as the relation between temporal 
and ensemble averages. 
\end{abstract}

\maketitle

\section{Introduction} 
\label{introduction}

The so-called ``sheet model'' is an interesting toy model
for the study of self-gravitating systems, or more generally
of systems with long-range interactions. It is simply the
one dimensional (1D) generalisation of Newtonian gravity,
consisting of particles interacting by attractive forces 
independent of their separation (or, equivalently, infinite
parallel planes embedded in three dimensions interacting 
via Newtonian  gravity). Because the particle trajectories are 
exactly integrable between crossings, it has the nice feature 
that its numerical integration can be performed with an 
accuracy limited only by machine precision. It has 
been the subject of (mostly numerical) study in the 
literature for several decades (see, e.g., 
\cite{yawn+miller_2003} for a review of the literature
on the model) following earlier analytical studies \cite{camm, oort_1932}.
A fundamental question about this system --- and more
generally for any system with long-range interactions ---
is whether they relax to the statistical equilibrium calculated 
in the microcanonical or canonical ensemble. For this
model the latter were first calculated exactly,  for any particle 
number $N$,  by Rybicki \cite{rybicki}.  The literature on this model --- which 
we will discuss in  greater detail in our conclusions section below ---  is marked by 
differing results (or, rather, interpretation of results) from different groups, 
and even some controversy.  Work by two groups in the eighties (see, e.g. 
\cite{reidl+miller_1987} for a summary) led to the conclusion
that relaxation could not be observed, except perhaps in
some special cases.  Studies by two other groups over a decade
ago  \cite{tsuchiya+gouda+konishi_1996, milanovic+posch+thirring}
found results indicating relaxation, and  \cite{tsuchiya+gouda+konishi_1996}
gave a  determination of the $N$ dependence of the characteristic 
time. However doubts about the interpretation of these latter
results as establishing relaxation to equilibrium
were raised by further analysis 
\cite{tsuchiya+gouda+konishi_1997, yawn+miller_erg_1997}.
In more recent work \cite{yawn+miller_2mass_1997, yawn+miller_2003}
clear evidence for relaxation in a version of the model in which
there are different particle masses has been found, but the
dependence on $N$ has not been determined\footnote{Other variants of
the model have also been studied 
in \cite{miller+youngkins_concentric_1998,valageasOSC_1,valageasOSC_2}.}
The mechanism of relaxation (if it indeed takes place) in these models 
remains, as in other long-range interacting systems, very poorly 
understood, and a basic subject of research in the statistical mechanics 
of long-range interacting systems (for recent reviews see 
e.g. \cite{campa_etal_LRreview_2009, bouchet+mukhamel_LRreview_2010}). 
In this article we report an essentially numerical study of relaxation in the 
single mass sheet model.
We introduce a simple but, as we will see, very useful tool 
for the characterisation of the long-time evolution and relaxation
of the system. This tool allows us to resolve some outstanding
issues about the relaxation in this system, and, in particular,
to establish more definitively both that relaxation does indeed 
occur and the scaling with particle number of the time 
characterizing it. We consider a broader range of initial 
conditions, which allows us to establish also dependences
of relaxation on these.  We also study the fluctuations --- both in 
time and over realizations of the initial conditions ---  about the 
average macroscopic evolution of the system,  showing
phenomenologically  the correlation of their amplitude with the
lifetimes of the intermediate ``quasi-stationary" states. 

We will discuss in greater detail in our conclusions the 
relation of our results to those in the previous literature, 
but it is useful at the outset to say a little more about the
more general context of this study. In recent years there has been
considerable interest in the statistical mechanics of long range
interactions (see, e.g., \cite{dauxoisetal, Assisi, campa_etal_LRreview_2009}),
stimulated by the need to understand the physics of 
various laboratory systems with interactions of this kind,
as well as by the more classical case of self-gravitating
matter relevant in astrophysics and  cosmology.
In this context one toy model in particular, and various
variants of it, has been much studied:   the Hamiltonian Mean 
Field (HMF) model (see e.g. \cite{yamaguchi_etal_04, campa_etal_2007,chavanis+campa_HMF_2010} 
and references therein), which is a model of particles
on a circle interacting by a cosine potential.  Its study
has shown that it shares many of the qualitative features
well documented in the most studied of realistic long-range
interacting systems --- self-gravitating systems in astrophysics ---
and believed to be generic in such systems. Starting from generic 
initial conditions, the system evolves rapidly (by ``violent relaxation'')
to a virialized macroscopically stationary state. These
states  --- commonly
referred to in the more recent literature as ``quasi-stationary
states'' (QSS) --- are out of equilibrium states, which are
described theoretically in the framework of Vlasov equation
(more usually referred to as the ``collisionless Boltzmann equations''
in the astrophysical literature). On much longer time scales an 
evolution towards the true thermal equilibrium (i.e. that determined 
by the maximization of the Boltzmann entropy in a mean field
approximation) is postulated.  For realistic systems --- such as
Newtonian gravity in three dimensions ----  it is very difficult 
numerically to simulate the evolution on sufficiently long
time scales to probe the relaxation. Studies in the literature 
(see e.g. \cite{theis+spurzem_1999, diemandetal_2body, knebe, elzant_2006, levin_etal_2008})
provide some results but give still a very limited characterization 
and understanding of it.  The HMF model
has the particular feature that the potential energy of any particle 
may be expressed as a function of its (angular) position and the 
mean potential energy due to all particles --- it is for this
reason that it is ``mean-field'' ---   so that the calculation of the forces 
in a system with $N$  particles requires only of order $N$ 
operations (rather than $N^2$ in a typical long-range interacting 
system).  Further the force is continuous at zero separation, 
so that the difficulties associated in the case of gravity with 
the regulation of the potential at small scales are avoided.
This allows the regime of relaxation to be accessed numerically
even for quite large particle numbers.  The study of 
 \cite{yamaguchi_etal_04} found a scaling of the relaxation 
 time in proportion to $N^{1,7}$ (but see also e.g. \cite{campa_etal_2007}
 which finds indications of longer lifetimes for other initial conditions). 
 
It can clearly be of interest to study different toy models, to
determine in particular features which are indeed generic.
The ``sheet model" is probably the oldest toy model of
long range interactions --- it was first explored in astrophysics
as a toy model for self-gravitating systems  in three dimensions ---
and is also, arguably, closer to reality than the HMF which
is constrained on a circle. It has, further, as mentioned
above the nice feature that it numerical integration can be
performed up to machine precision. Despite this, the results 
concerning its dynamics and relaxation are less clearly 
determined than for the HMF, and the literature on the
subject has, as we have discussed above,  been marked by 
some controversy and results showing that the model has, apparently, some
very peculiar behaviours --- rapid relaxation to equilibrium
for some classes of initial states \cite{luwel+severne+rousseeuw_1983, reidl+miller_1991}, persistent phase 
space structures impeding relaxation to QSS \cite{rouet+feix, mineau+rouet+feix_1990}, macroscopically 
chaotic behaviour in the long time evolution \cite{tsuchiya+gouda_2000} --- which 
indicate that it might not be a very useful toy model (in that its
behaviours are perhaps non-generic). In this article our main
conclusion is that, 1) by using appropriate diagnostics of the 
macroscopic evolution and 2) by  extending simulations
to sufficiently large $N$ and/or averaging over sufficiently
large numbers of realization,  one finds behaviour in 
this toy model very similar in crucial respects to that in the 
HMF:  to a very good first approximation a generic initial
configuration relaxes to a long-lived QSS, and
then relaxes to its statistical equilibrium at sufficiently 
long time. This latter phase can be characterized 
apparently by a single time-scale, with the evolution
of the order parameter during relaxation well fit by a 
simple function (in our case a better fit is obtained 
using a simple stretched exponential, rather than 
a hyperbolic tangent in the HMF as in \cite{yamaguchi_etal_04}).
On the other hand the $N$ dependence of this time
scale,  linearly proportional to the number of particles $N$,
is different to that found in  \cite{yamaguchi_etal_04} for the
HMF. This latter result, however, applies to spatially
homogeneous states which, in the HMF, can occur
due to the periodicity of the system.  Relaxation which
is slower than linear in $N$ is expected in this case, as
shown using  analysis of kinetic equations (see, e.g.
contributions of P.H. Chavanis, and of  F. Bouchet 
and J. Barr\'e in \cite{Assisi}).

The article is organised as follows. In the next section 
we recall the basic definitions of the model, and relevant 
results on its statistical equilibrium.  We then explain the choice 
of the macroscopic parameters  (``order parameters'') we choose
 to monitor the evolution of  the system. In the following section
 we first describe our numerical simulations and the initial conditions 
 we study, and then give our results. In presenting them
 we give first results for single realizations, and then use
 temporal averages and finally ensemble averages to derive
 the scaling with $N$ of the relaxation time. This is followed
 by further study of the fluctuations about the average
 behaviours of the order parameters.  Considering 
 both temporal fluctuations and those in the ensemble,
 which we show to be very consistent with one another, 
 we observe the correlation between their amplitude in
 the QSS and the observed relaxation time. In the
 conclusion sections we return to a more detailed
 discussion of the previous literature, presenting
 further results which allow one to understand the
 reasons for the divergence in conclusions in
 certain cases.

\section{The sheet model} 
\label{basic_definitions}

We first recall the model and fix our notation. We next 
summarize the results of \cite{rybicki}
on statistical equilibrium, and then explain the rationale
for our choice of ``order parameter'' in our study.

\subsection{Definitions} \label{definitions}

We consider identical (equal mass) point particles in one 
dimension
interacting by an attractive force independent of 
separation, i.e., the 
force $f(x)$ on a particle at coordinate 
position $x$ exerted by a particle at the origin is given by 
\begin{equation}
\label{Familiar_pair_force}
 f_{1d} (x) = -g\frac{x}{\vert x \vert} = -g ~\textrm{sgn}(x)~,
\end{equation}
where $g$ is the coupling. Equivalently it is the pair
interaction derived from the pair potential 
\begin{equation}
\psi(x) = g|x|
\label{pair_potential}
\end{equation}
which satisfies the 1D Poisson equation
for a point source, $\frac{d^{2}\psi}{dx^{2}}=2g \delta_{D} (x)$
(where $\delta_D$ is the Dirac delta function). Comparing with 
the three dimensional (3D) Poisson equation
shows the equivalence with the case of an infinitely 
thin plane of infinite extent and surface mass density 
$\Sigma=g/2\pi G$, which explains the widely used
name ``sheet model''. We will work in the one
dimensional language, referring to ``particles''. 
For a particle at coordinate position $x$ on the real
axis the total force $F(x)$ acting on it is thus
\begin{equation}
  F(x) = g \Big[ N_{>}(x)- N_{<}(x) \Big].
\label{force-finite-system}
\end{equation}
where $ N_{>}(x)$ and $N_{<}(x)$ are, respectively, the number
of particles with coordinates greater than or less than $x$ 
(i.e. the force on a given particle is proportional to the 
difference in the number of particles on its right and 
its left). 

To specify fully the dynamics we must prescribe what happens 
when two particles arrive at the same point. Since the force is
bounded as the separation goes to zero, the natural physical 
prescription for the 1D model is that the particles simply
cross (i.e. pass through one another). In one dimension, however, 
this is equivalent, up to a change in particle labels, to
a hard elastic collision, as such a collision (of equal
mass particles) simply results in an exchange of their 
velocities. Thus, up to particle labels, the sheet model 
for equal masses is equivalent to one in which particles
experience always the same spatially constant force  
Eq.~(\ref{force-finite-system}) and simply exchange 
velocities when they ``collide''. As has been noted 
in some previous studies of models of this kind
\cite{aurell_etal} it is convenient
to exploit this equivalence in numerical simulation,
as will be described below.

\par 

In contrast to Newtonian gravity in three dimensions, the pair potential (\ref{pair_potential}) is 
positive and diverges at large separations, so that particles cannot escape from the system to
infinity. It has therefore no particular interest to enclose the system in a 
finite box, and indeed such a confinement is not necessary in order to 
define the statistical equilibrium (in contrast to three dimensions). 
We will consider therefore always open boundary conditions. Likewise 
the fact that the potential  has no divergence at short distances means 
that there is no equivalent of the so-called ``gravo-thermal collapse''
in three dimensions.

\subsection{Thermal equilibrium}
\label{te}
It has been shown by Rybicki \cite{rybicki} that the statistical equilibrium for 
this model 
can be derived exactly in the microcanonanical ensemble, for any $N$. We will study here the
$N \rightarrow \infty$ limit (at fixed energy and mass, see~\cite{rybicki} 
for full derivation), in which 
case the phase space distribution function (i.e. mass per unit phase
space volume) becomes: 
\begin{equation}
{f}_{\rm eq} (x,v)=\frac{M}{ 2 \sqrt{\pi} \sigma\Lambda} \, e^{-\frac{v^{2}}{\sigma^{2}}} \, \textrm{sech}^{2}\frac{x}{\Lambda}
\label{rybicki-eq}
\end{equation}
where $\sigma$ and $\Lambda$ are the characteristic scales of velocity 
and length, and $M$ is the total mass of the system.
It is straightforward to verify that
\bea
\sigma^{2} &=& \frac{4E}{3M} \label{sigma} \\
\Lambda &=& \frac{4E}{3gM^{2}}. \label{lambda}
\eea
where $E$ is the total energy of the system, which allows
one to calculate ${f}_{\rm eq} (x,v)$ explicitly as a function
of $M$ and $E$ (and $g$) only.

As is typical of long-range systems, the statistical equilibrium
is thus characterised by a space independent Maxwellian velocity 
distribution and an inhomogeneous spatial distribution. The same
solution is recovered in the canonical 
ensemble. Thus, differently to many long-range systems (including 
gravity in three dimensions)
the two ensembles are completely equivalent. This behaviour is associated 
also with
the absence of microcanonical phase transitions which may arise in such systems. 

This equilibrium solution in the continuum limit can be most easily 
derived by simply maximizing the Boltzmann entropy at fixed mass and 
energy, using the mean-field expression for the energy:
\bea
E=\frac{1}{2} \int v^2 f(x,v) dx dv + 
 \frac{1}{2} \int f(x,v) \psi (|x-x'|) f(x',v') dx dx' dv dv' \,.
\eea
This procedure gives simply 
\be
{f}_{\rm eq} (x,v)=A e^{-\beta [\phi(x) + \frac{v^{2}}{2}]}=
\frac{\pi^{-1/2}}{2}e^{-\frac{v^{2}}{\sigma^{2}}} \rho(x)
\label{boltzmann_eq}
\ee
where $\phi(x)$ is the mean field potential
\be
\phi(x) =  \int \psi (|x-x'|) f(x',v') dx' dv'\,.
\ee
and  $\rho(x)$ the associated mass density profile,
which is therefore the solution to 
\be
\frac{d^{2}}{dx^{2}} [\ln \rho(x) ]= -2 g \beta \rho (x).
\ee
It is simple to verify that the expression Eq.~(\ref{rybicki-eq}) 
results, with appropriate identification of constants
and choice of units.

\subsection{Order parameters for relaxation} 
\label{op}

To monitor relaxation to equilibrium it is possible in principle to
simply study the full distribution function as a function of time. In practice
relaxation is extremely slow (in the characteristic time units of the system)
and only accessible numerically for relatively small numbers (of order a 
thousand) particles, which makes the comparison of the full function
subtle (because of finite $N$ fluctuations). In the previous literature 
various methods have been used --- statistical tests 
on the velocity and spatial distributions (e.g. 
\cite{luwel+severne+rousseeuw_1983, reidl+miller_1987}) 
and analyses based on
the evolution  of particle energies coarse-grained in time
\cite{tsuchiya+gouda+konishi_1996}).  Here instead we  
study the evolution primarily using appropriately
chosen macroscopic parameters, i.e., ``order parameters''  which take, 
in general, distinct values 
in and out of equilibrium. This is somewhat analogous to the 
approach used in the study of the HMF, where the magnetisation of 
the system plays the role 
of an order parameter used to characterize the evolution out
of equilibrium (see e.g. ~\cite{yamaguchi_etal_04}). 
Once the expected behaviour of the macroscopic  
parameter is identified, a more detailed analysis involving the 
distribution functions can be used to confirm that the system 
has indeed fully relaxed. We will see that, with the choice of
parameter we make, it turns out that the single macroscopic 
parameter is sometimes a better indicator of relaxation that
the full density or velocity distributions, and that indeed some
of the controversial results in the previous literature may
easily be sorted out using the tools used here.

An evident property of the distribution function Eq.~(\ref{rybicki-eq})
is that it is separable in its spatial and velocity coordinates.
It is simple to show, as we will now verify, that it is in fact also
the unique stationary solution of the Vlasov equation which is
separable. Thus if the system is, during its evolution, very close
at any time to a stationary solution of the Vlasov equation (which
describes the collisionless limit) any parameter probing the 
degree of separability of the distribution function would be
expected to be a useful indicative measure. This leads us to
consider order parameters which are simply suitably 
normalized moments of the distribution function. 

Let us first verify the result on separability. The Vlasov equation for 
the model is
\begin{equation} 
\frac{\partial f}{\partial t}+v\cdot\frac{\partial f}{\partial x}+ a(x)\cdot\frac{\partial f}{\partial v}=0.
\label{vlasov}
\end{equation}
where $a(x)$ is the mean field acceleration, i.e., 
\begin{eqnarray}
a(x)&=& g \int {\rm sgn} (x'-x) f(x',v) dx' dv \nonumber \\
&=& g \int {\rm sgn} (x'-x) \rho(x') dx' 
\label{vlasov_a}
\end{eqnarray}
which can be conveniently expressed in terms of the
mean field potential $\phi(x)$ satisfying the 
Poisson equation with $\rho(x)$ as source , i.e., 
\be
\frac{d a(x)}{dx}=-\frac{d^{2}\phi}{dx^{2}}= -2g \rho(x)
\label{poisson-mean-field}
\ee
with an appropriate boundary condition from 
Eq.~(\ref{vlasov_a}) [e.g. $a(x \rightarrow +\infty) = gM$,
where $M$ is the total mass].
Seeking a solution which is both {\it stationary and separable} we take
\begin{equation}
f(x,v,t) \equiv \rho (x) \theta (v)
\label{separated_f_vlasov}
\end{equation}
and thus obtain, on substitution,
\begin{equation}
\theta (v)\cdot v\cdot\frac{\partial \rho (x)}{\partial x}=-\rho (x)\cdot a(x)\cdot\frac{\partial \theta(v)}{\partial v}\,.
\label{stationary_vlasov}
\end{equation}
Given that $a(x) \neq 0$ everywhere (except at the single point 
which divides the mass in two) we can
write this as \footnote{Note that this
is not true in the HMF model, as the magnetization 
(which determines the acceleration) can indeed be zero everywhere
when the stationary state is spatially uniform.  This is a result 
of the periodic nature of the system. In this case there may thus 
exist QSS which are separable, uniform in space but with a non-
maxwellian velocity distribution. Such QSS are indeed observed
and have been extensively studied in this model
(see e.g. \cite{yamaguchi_etal_04}).}
\begin{equation}
-\frac{1}{v\cdot \theta (v)}\cdot\frac{\partial \theta(v)}{\partial v}=
\frac{1}{a(x)\cdot\rho (x)}\cdot\frac{\partial \rho (x)}{\partial x}.
\label{vlasov_side_by_side}
\end{equation}
for any region of $x$ where $\rho(x) \neq 0$. It follows that 
both sides are equal to a constant, $C$ say, and therefore 
\begin{equation}
\theta (v) = \theta_{0}e^{-\frac{C}{2}v^{2}}.
\label{vlasov_theta}
\end{equation}
The right hand side gives 
\begin{equation}
\frac{1}{\rho (x)}\cdot\frac{\partial \rho (x)}{\partial x} = C a(x)\,.
\label{eq-separated-patial}
\end{equation}
Differentiating with respect to $x$ and using the Poisson
equation for the mean field Eq.~(\ref{poisson-mean-field}), this 
gives exactly the same equation used above to determine 
the equilibrium solution for $\rho(x)$. The expression Eq.~(\ref{rybicki-eq}) is
indeed therefore the only stationary separable solution 
of the Vlasov equation.

Given this observation we define the following order parameters:
\begin{equation} 
\phi_{\alpha \beta} = \frac{\langle|x|^{\alpha}|v|^{\beta}\rangle}{\langle|x|^{\alpha}\rangle \langle |v|^{\beta}\rangle}-1 
\label{phi}
\end{equation}
for non-zero $\alpha$ and $\beta$, where 
\begin{equation} 
\langle u \rangle \equiv \frac{1}{N} \sum_{i=1}^{N} u_{i}
\label{phi_def}
\end{equation}
and $u_i$ is the value of the parameter $u$ for the $i$-th particle. 
By construction these quantities are zero in thermal equilibrium.
While a finite number of such moments can of course be zero in
a QSS with a non-separable distribution function, we expect them
generically to be non-zero in such states. We will use here 
both  $\phi_{11}$ and $\phi_{22}$. As detailed in the next section,
we will consider both their temporal evolution in single realizations 
of our initial conditions, as well as averages of these temporal
evolutions. These averages will be performed in two different
ways:  by averaging over a finite temporal window in a single 
realization,  and by averaging over independent realizations of 
the initial conditions. Further we will consider the evolution 
as a function of time of the fluctuations of $\phi_{11}$ and 
$\phi_{22}$ with respect to these averages. 

\section{Numerical simulations} \label{simulations}

\subsection{Algorithm}
As remarked above in Sec.~\ref{definitions}, it is convenient for the numerical integration 
of the model to exchange particles' labels when they cross, which is equivalent 
to treating them as if they undergo an elastic collision in which they 
exchange their velocities when they meet. 
The force on each particle is then constant in space and time
[and given by Eq.~(\ref{force-finite-system})], 
and the numerical algorithm must simply determine, at any time, the next crossing which occurs, 
and then exchange the velocities of the ``colliding'' particles. The optimal way to treat this 
kind of problem is, as has been pointed out and discussed in detail in \cite{noullez_etal},
by using a so-called ``heap-based'' algorithm, which uses an object
called a ``heap'' to store in an ordered way the next crossing 
times of the pairs (see \cite{noullez_etal} for details). 
This algorithm requires a number of operations of order $\log(N)$ to 
determine the next crossing (rather than of order $N$ for the evident 
direct algorithm in which one calculates and compares directly at each step
the next crossing of each of the $N-1$ pairs). Given that the number of 
crossings per particle per unit time grows in proportion to $N$, the 
simulation time thus grows in proportion to $N^{2}log(N)$.

Because the particle trajectories are integrated exactly, the only limit on the accuracy of the 
numerical integration is thus the numerical precision. As is common practice we will use the
total energy (which is conserved in the continuum model) as a control parameter. For the
longest simulations we report the error in total energy of the order of $10^{-8}\%$. 
\subsection{Initial conditions}

\begin{figure}[h!]
\begin{center}
  \includegraphics[width=8cm]{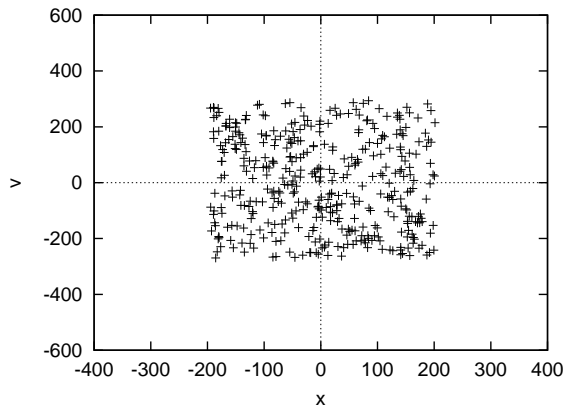}
\caption{A rectangular waterbag initial condition in phase space sampled
with $N=400$ particles, for a initial virial $R_{0}=1$ (see text for definition of units).} 
\label{ic_example}
\end{center}
\end{figure}

We will consider principally a simple class of spatially uniform initial conditions (IC), 
generated by randomly distributing the $N$ particles on a finite interval. As initial velocity
distribution we will consider both the case that initial velocities are zero
(``cold IC'') and the case that this distribution is also uniform in a finite interval. 
The latter are thus random samplings of a particular class  of ``waterbag'' initial 
conditions in phase space (i.e. in which the phase space density in equal in
the region in which it is non-zero), while the cold case can be considered 
as the limit in which the width of the velocity distribution goes to
zero. In Fig.~\ref{ic_example} the phase space distribution 
for a typical IC is shown. 

These IC may be characterized solely by the particle number $N$ and 
a single parameter characterizing the waterbag. 
Rather than the width of the velocity distribution or phase space density, it is 
convenient to choose the parameter characterizing the waterbag to be the 
dimensionless {\it initial virial ratio}: 
\begin{equation}
R_{0}\equiv \frac{2T_{0}}{U_{0}} 
\label{virial_ratio}
\end{equation}
where $T_{0}$ is the initial average total kinetic energy and $U_{0}$ 
is the initial average total potential energy. By ``average" here we
mean that the values of  $T_{0}$ and  $U_{0}$  are those
calculated for the theoretical waterbag configuration. When we 
consider, as we do below,  different realizations of the initial conditions 
at fixed $N$ and $R_0$ there are of course finite $N$ fluctuations 
about these values of $T_{0}$ and  $U_{0}$. The latter correspond
then to their average values in the ensemble 
of initial conditions at fixed $N$ and $R_0$. We note that this ensemble 
of initial conditions is clearly  {\it not} a subset of the microcanonical 
ensemble because there are finite $N$ fluctuations also about the average total
energy. As these fluctuations are small, however, we will assume
below that the evolution of such an ensemble of initial conditions 
represents well that of an ensemble of such initial conditions 
with exactly the average energy\footnote{It is simple to 
show, given that the particle positions and velocities are both 
randomly sampled from a PDF which is uniform in a finite 
interval, that the relative fluctuations in $U_0$ and $T_0$ scale 
as $1/\sqrt{N}$ for large $N$.  An exact calculation shows, for 
example, that at $N=100$, the normalized variance of  $U_0$, which 
corresponds to that in the energy for the case $R_0=0$,  is $\approx 0.05$. 
This means that the typical amplitude of the fluctuation in the energy for cold IC 
at $N=100$ is of order the difference between the mean energy
of cold IC and IC with $R_0=0.1$.}.

We remark on a particularity of the cold IC which we will return
to below. In the limit $N \rightarrow \infty$, the evolution
from this IC becomes singular at a finite time: an element
of mass initially at coordinate position $x_0$ feels a 
force $-2 g \rho_0 x_0$, where $\rho_0$ is the initial 
mass density; all particles are in free-fall under a 
force proportional to their distance, and therefore
arrive at the origin at the same time, producing a 
density singularity. For a finite system, the corresponding behaviour
is associated to the existence of a periodic oscillating mode
when the particles are initially equally spaced (i.e. on a
regular lattice). This ``breathing mode'' of such a cold
system has been discussed in \cite{reidl+miller_1993}.
While there is no such mode in a three dimensional
system of a finite number of particles,  the $N \rightarrow \infty$
limit of a cold spherical initial condition has the analogous
singularity (see \cite{ejection_mjbmfsl} for a detailed discussion).

\subsection{Units and coordinates}

For convenience we choose our coordinate system such that the centre of the mass of the
system lies at $x=0$ and is at rest (i.e. after distributing the particles as described 
we add a spatial translation and constant velocity to all particles so that these 
conditions are satisfied). 

We make the following choice of units: we set the particle mass $m$ and the coupling $g$ to
unity, and take $L=N$. This corresponds to a mass (and particle) density of unity, and
a time unit equal to the {\it dynamical time}:
\be
t_{\rm dyn}= \sqrt{\frac{Lm}{Ng}}
\ee 
which is the characteristic time for the system's evolution under the 
mean field forces (the mean field forces, of order $Ng$, moves a system 
particle of mass $m$ over the system size $L$ on this time-scale).
$t_{\rm dyn}$ also coincides with the time of the singularity
noted above in the smooth limit of the cold IC.

\section{Results} \label{results}

The difficulty in this study of relaxation, as in such a study for any
long-range system, is that one is interested in studying large $N$ 
systems --- so that finite $N$ deviations from the mean-field 
behaviour are small --- on a time scale which grows rapidly 
with $N$ (typically, one expects, in proportion to some power of 
$N$). Because of numerical limitations, particularly strong because of
the computational cost of integrating a long-range  interaction, it is in 
practice often difficult to arrive at definitive conclusions.  In the
case of gravity in three dimensions, notably, numerical studies
exist (see references above) but they give only a very incomplete
characterization and understanding of relaxation.
As we have discussed
in the introduction one of the attractive features of the HMF model is 
that, because of its mean field nature, the numerical cost of the force 
calculation is of order $N$, allowing much larger particle numbers
--- $N \sim 10^4-10^5$ \cite{yamaguchi_etal_04} --- to be simulated 
on the relevant long time scales than is feasible in other cases.
The  principal reason why the early literature on the sheet model was 
marked by controversy on the question of relaxation is simply, as we
will discuss further below,  that such relaxation could not be 
observed on the required time scales for systems sufficiently large 
so that the finite $N$  fluctuations were sufficiently small to allow 
the clear identification of the average behaviours. The study of
\cite{tsuchiya+gouda+konishi_1996}, taking advantage of the
greater numerical resources available already in the nineties,
detected relaxation for $N\sim 10^2$ from specific waterbag
configurations and found a scaling of the relaxation time, over
a small range in $N$,  linear in $N$. This result was obtained, 
however, by doing a time average of their chosen diagnostic 
over a very broad time window (of order $10^5$, only an
order of magnitude less than the typical relaxation time for
the cases explored), and its solidity has been placed in
question in subsequent work \cite{tsuchiya+gouda+konishi_1997,yawn+miller_erg_1997}.
Exploiting the increase in numerical power since then, 
and aided greatly by the diagnostics we have defined in the previous
section, we report here
\footnote{Evolution of $N=10^2$ particles to $t=10^6$ 
requires about $20$ minutes on a single processor; thus,
given the scaling with $N^2 \log N$ of the computational
cost per unit physical time, and a linear growth (see
below) in the relaxation time itself, simulation
times of order several weeks are required for the
most rapidly relaxing case with $N=10^3$. Our largest
$N$ results are ensemble averages over systems
with $N=800$, obtained by running simultaneously
on a large number of work stations.} results showing relaxation 
for systems with $N\sim 10^3$. Further we obtain 
our results for the scaling of the relaxation time by 
doing ensemble averages (over realizations of the
initial conditions) without time averages.

\subsection{Temporal evolution of order parameters} 
\label{relaxation}

\begin{figure}[h!]
\begin{center}
\begin{tabular}{ccc}
  \includegraphics[width=5cm]{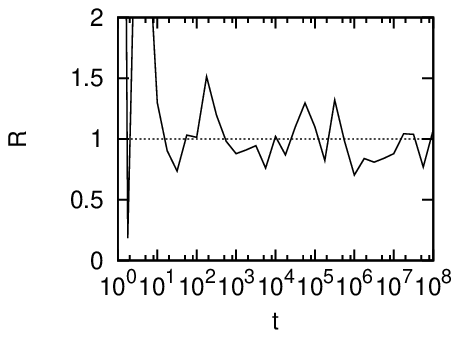} & \includegraphics[width=5cm]{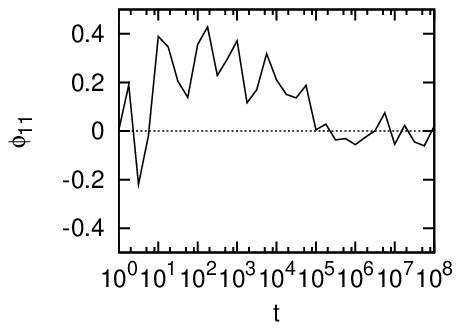} 
& \includegraphics[width=5cm]{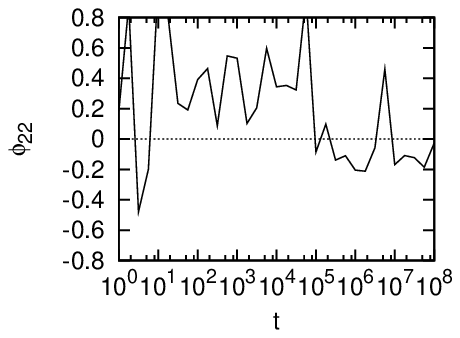}
\end{tabular}
\caption{Evolution as a function of time of the virial ratio (left panel), $\phi_{11}$ (middle panel) and $\phi_{22}$ (right panel),
in a simulation with $N=100$ particles and initial virial ratio $R_{0}=0$. We observe that the system virializes on a time
scale of a few tens of dynamical times; further  the behaviours of $\phi_{11}$ and $\phi_{22}$ indicate, as typically expected
in long-range interacting systems, a subsequent evolution in a long-lived QSS which eventually relaxes to thermal 
equilibrium  (in which  $\phi_{11}=0=\phi_{22}=0$ on average).} 
\label{fig_a}
\end{center}
\end{figure}

\begin{figure}[h!]
\begin{center}
\begin{tabular}{ccc}
  \includegraphics[width=5cm]{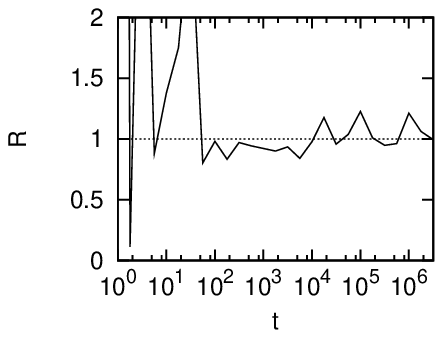} & \includegraphics[width=5cm]{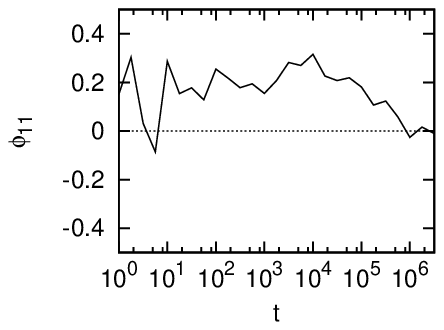} 
& \includegraphics[width=5cm]{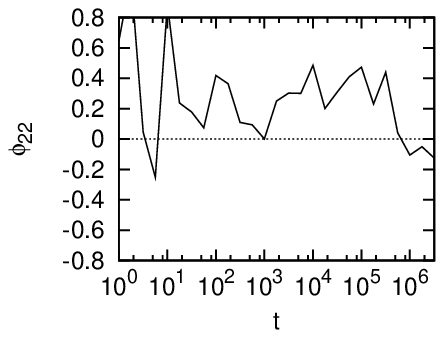} 
\end{tabular}
\caption{Evolution as a function of time of the same three parameters as in Fig. ~\ref{fig_a}, in a simulation 
with $N=400$ and initial virial ratio $R_{0}=0$. We see the same qualitative behaviours as in the previous
figure, except that fluctuations are of lower amplitude and the QSS phase appears to persist longer.} 
\label{fig_b}
\end{center}
\end{figure}

Shown in Fig.~\ref{fig_a}  is the evolution of the
virial ratio $R$, and the order parameters  
$\phi_{11}$ and $\phi_{22}$ in a {\it single realization} of
a system with $N=100$, and $R_0=0$. Note that the time
axis (as it will be invariably here) is logarithmic.
In   Fig.~\ref{fig_b} are plotted the same 
quantities for $N=400$.  While the fluctuations are very
large, particularly in the first case, one can make
out that there are, as expected, two stages in the
macroscopic evolution probed by these parameters: a 
first stage ($ t < 100$) of ``violent relaxation'' 
during which all quantities (and notably the virial
ratio) fluctuates strongly before settling down to
behaviours which appear to fluctuate about a well
defined average, and specifically about unity for
the virial ratio. The averages of the parameters
$\phi_{11}$ and $\phi_{22}$ are clearly non-zero
on a much longer time scale than that characterizing
the virialization. These non-zero averages, which
appear to be approximately the same in each case 
for the two different $N$, appear to remain roughly stable until at least
about $10^4-10^5$, after which both $\phi_{11}$ and 
$\phi_{22}$ start to evolve towards zero.  The
time scale at which the evolution sets in is
clearly significantly shorter for $N=100$. This
behaviour should indicate, as we have discussed
above, the relaxation to statistical equilibrium.

\begin{figure}[h!]
\begin{center}
\begin{tabular}{ccc}
  \includegraphics[width=5cm]{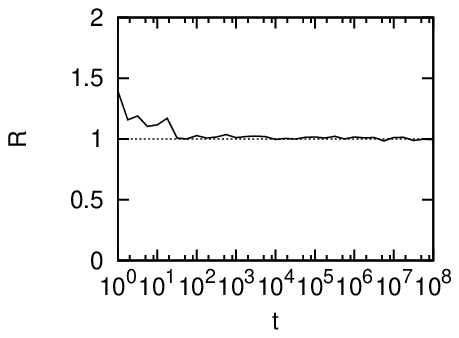} & \includegraphics[width=5cm]{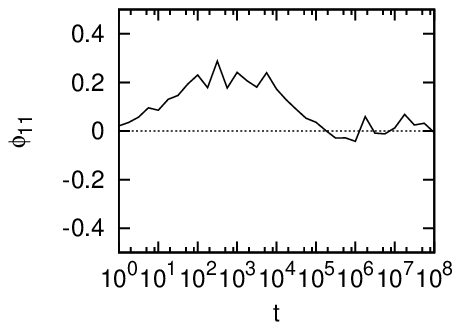} 
& \includegraphics[width=5cm]{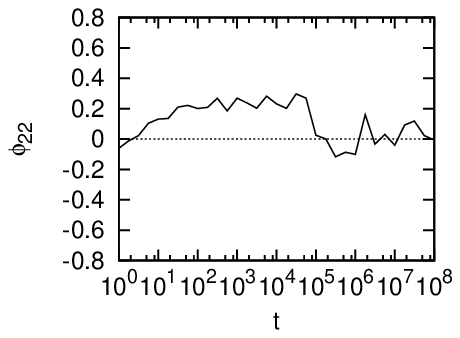} 
\end{tabular}
\caption{Evolution as a function of time of the same three quantities and the same
initial conditions as in Fig. ~\ref{fig_a}, but with these quantities now averaged in 
a time window of width $\Delta t=10$ as described in text. The averaging makes
the interpretation given in Fig. ~\ref{fig_a} much clearer: once virialized the
system stays in a long-lived QSS and eventually relaxes to thermal equilibrium. } 
\label{fig_c}
\end{center}
\end{figure}

\begin{figure}[h!]
\begin{center}
\begin{tabular}{ccc}
  \includegraphics[width=5cm]{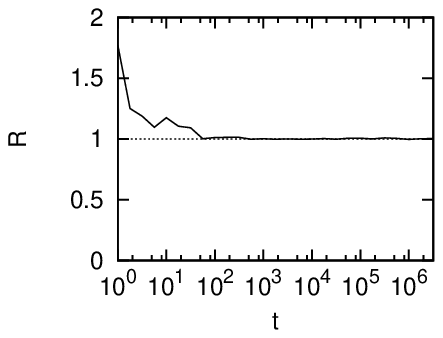} & \includegraphics[width=5cm]{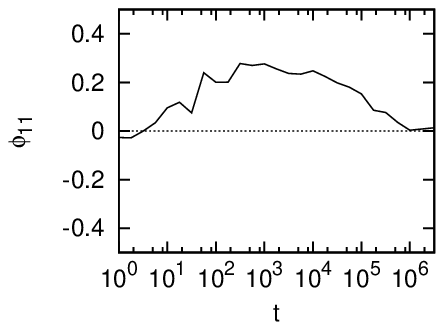} 
& \includegraphics[width=5cm]{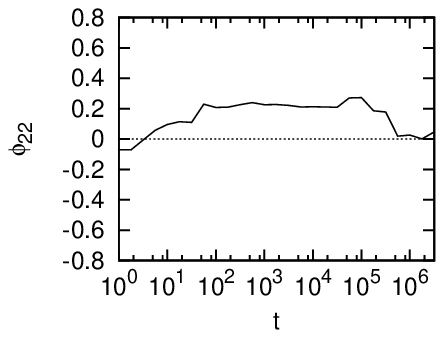} 
\end{tabular}
\caption{Evolution as a function of time of the same quantities and for the same
initial conditions with $N=400$ as in Fig.~ \ref{fig_b}, but with these quantities now averaged in 
a time window of width $\Delta t=10$. Comparing to the previous figure (same
quantities for $N=100$) we see clearly that the QSS persists for longer.} 
\label{fig_d}
\end{center}
\end{figure}

These behaviours can be seen more clearly by
averaging in a temporal window, of width small
compared to the characteristic times scales of
this apparent evolution. Shown in 
Fig.~\ref{fig_c} and \ref{fig_d} are the same 
quantities for the same simulations, but now each
point represents the average over one hundred
time slices, equally spaced in a window of width
$\Delta t=10$ centred on the given time.

\begin{figure}[h!]
\begin{center}
\begin{tabular}{cc}
  \includegraphics[width=7cm]{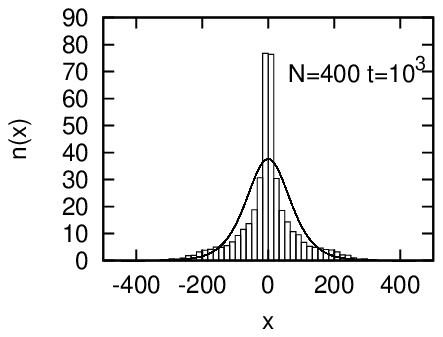} & \includegraphics[width=7cm]{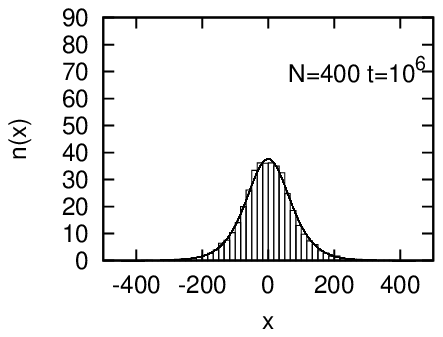} \\
  \includegraphics[width=7cm]{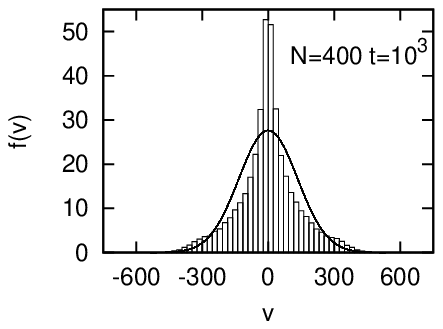} & \includegraphics[width=7cm]{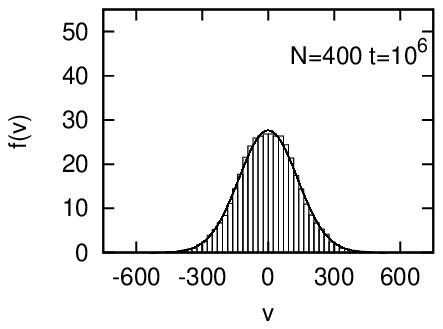}
\end{tabular}
\caption{Density profiles (top, left and right panels) and velocity distributions (bottom, left and right panels) 
at $t=10^{3}$ (left) and $t=10^{6}$ (right), in a simulation with $N=400$ particles started from 
initial conditions with $R_{0}=0$. The quantities are averaged in a temporal window of width $\Delta t$
exactly as in the previous two figures. The continuous lines are the expected distributions
at thermal equilibrium, Eq.~(\ref{rybicki-eq}).}
\label{fig_e}
\end{center}
\end{figure}

These behaviours are thus clearly indicative of the 
evolution expected, which is that believed to be typical of
long-range interacting systems: violent relaxation
brings one on a short time scale to a QSS, as a result
of a mean field dynamics described by the coupled
Vlasov-Poisson equations (and this independent of
$N$). On longer, $N-dependent$ times scales, one
relaxes to the mean-field equilibrium, given in this
case by Eq.~(\ref{rybicki-eq}). That the decay to zero of
$\phi_{11}$ and  $\phi_{22}$ does indeed correspond
to relaxation to the statistical equilibrium of Eq.~(\ref{rybicki-eq}) 
can be tested in further detail.  Fig. ~\ref{fig_e}
shows the velocity and space distributions
for $R_0=0$ and $N=400$ particles, averaged again over a 
time window of width $\Delta t=10$, at $t=10^3$ and  $t=10^6$.
The continuous lines correspond to Eq.~(\ref{rybicki-eq}),
clearly in very good agreement at the later time, and 
very different in the QSS phase. We have also 
checked (but do not show here) the agreement of the 
distribution of particle energies.  These results indicate
that $\phi_{11}$ and  $\phi_{22}$ are very good
diagnostics of the evolution towards equilibrium: indeed 
below  we will see that they are typically more discriminating 
of relaxation than the full density and velocity distributions.  

\subsection{Dependence on initial virial ratio} 

\begin{figure}[h!]
\begin{center}
\begin{tabular}{cc}
  \includegraphics[width=7.5cm]{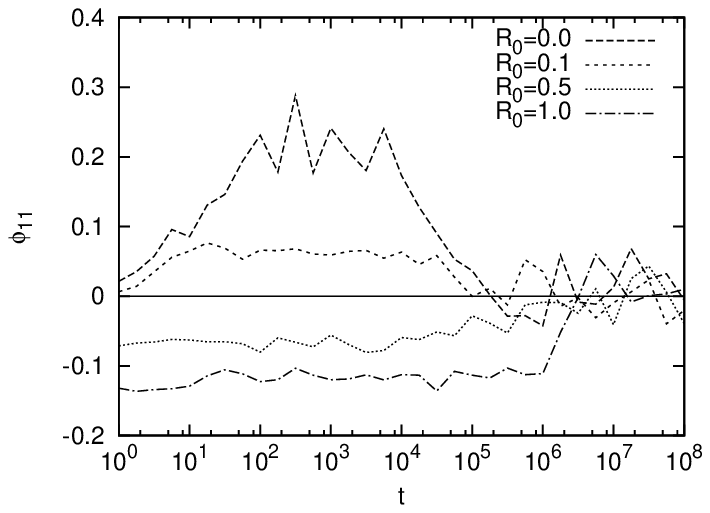} & \includegraphics[width=7.5cm]{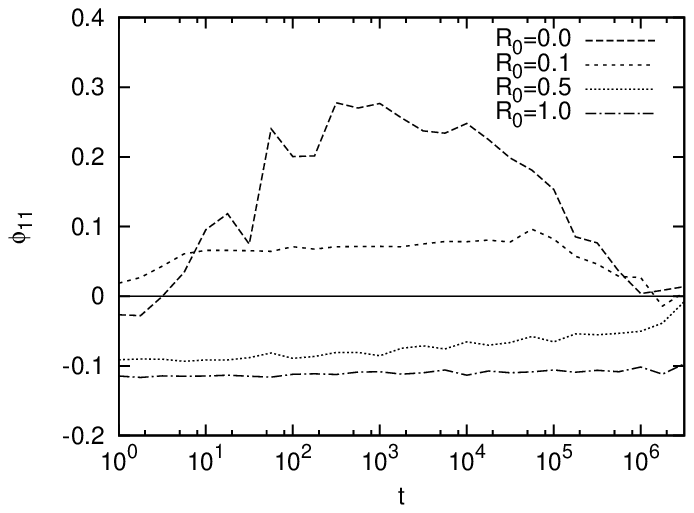} 
\end{tabular}
\caption{Evolution as a function of time of $\phi_{11}$,  averaged in a time
window of width $\Delta=10$, for  simulations with $N=100$ particles (left panel) and
$N=400$ particles (right panel), starting from initial conditions with the 
indicated values of $R_0$. The time scale for relaxation of a QSS clearly depends not just on 
$N$ but on the details of this state. }
\label{fig_g}
\end{center}
\end{figure}

\begin{figure}[h!]
\begin{center}
\begin{tabular}{cc}
  \includegraphics[width=7.5cm]{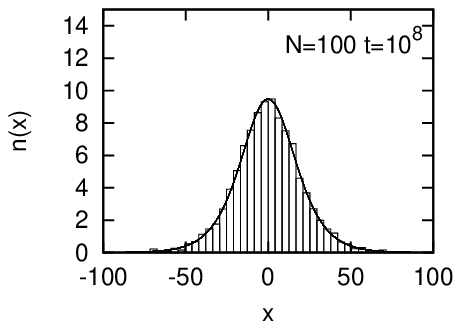} & \includegraphics[width=7.5cm]{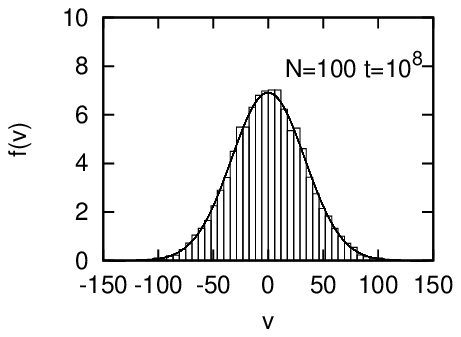} 
\end{tabular}
\caption{Density profile (left panel) and velocity distributions  (right panel),  averaged in a time
window of width $\Delta=10$ centered at $t=10^{8}$, for a simulation with $N=100$ particles
started from a virial ratio $R_{0}=0$. The continuous lines are the expected distributions
at thermal equilibrium, Eq.~(\ref{rybicki-eq}).}
\label{fig_f}
\end{center}
\end{figure}

Shown in Fig.~\ref{fig_g} are 
the evolution of $\phi_{11}$ for $N=100$ and
$N=400$ starting now from four different values 
of $R_0$, as indicated ($0, 0.1, 0.5, 1$) 
The results are averaged again in a time window 
of width $\Delta t=10$.  Note that results for $N=100$, which extend up 
to $t=10^8$, indicate that the evolution towards the
statistical equilibrium is really a relaxation to
a definitive equilibrium behaviour,  i.e.,  which
is stable and persists. This is further confirmed by
Fig. ~\ref{fig_f}  which shows the spatial and  velocity distributions 
for the case $R_0=0$  and $N=100$ at $t=10^8$ (with the 
same time averaging window as used above in 
Fig. ~\ref{fig_e}).  Fig.~\ref{fig_g} showv that there are, nevertheless, very
significant fluctuations in  $\phi_{11}$ 
and  $\phi_{22}$. These could indicate
significant macroscopic, but  stochastic, deviations 
from the equilibrium persisting over very significant 
times  (see \cite{ tsuchiya+gouda_2000}). We will 
present evidence below that they are, as one would 
expect, finite $N$ effects, with an amplitude which
decreases as $N$ increases.

We observe in Fig.~\ref{fig_g} that, as expected, the QSS 
resulting from violent relaxation is clearly 
different depending on the initial condition,
with very different values of $\phi_{11}$.
Further the relaxation toward equilibrium is
evident in most cases, but at a time which 
depends not only on $N$, but also on $R_0$ 
(or the intermediate QSS state).
More specifically, the smaller is $R_0$ 
the shorter is the lifetime. Indeed for 
$R_0=1$ we just see the onset of the 
relaxation for the case $N=100$, but do not
see it at all for $N=400$. For $N=100$
there is a difference of a  factor of about one hundred
in the time at which relaxation appears to becomes clearly 
visible in the cases $R_0=1$ and $R_0=0$.
In the respect we remark that earlier studies
have not considered this kind of cold initial
condition, in which relaxation occurs more 
rapidly.

\subsection{Estimation of $N$ dependence using ensemble average} 

Let us focus now on the $N$ dependence of the relaxation.
We wish to determine the scaling with $N$ of the characteristic
time for relaxation, at a fixed value of the initial virial
ratio. Given the very significant noise in the order 
parameters at the particle numbers we can simulate
numerically up to times on which relaxation 
occurs to do so  we must average out these fluctuations. This can be
done using either a time average on a single realization (as above) 
or an average over realizations (or possibly some 
combination of both).

\begin{figure}[h!]
\begin{center}
\begin{tabular}{cc}
  \includegraphics[width=7cm]{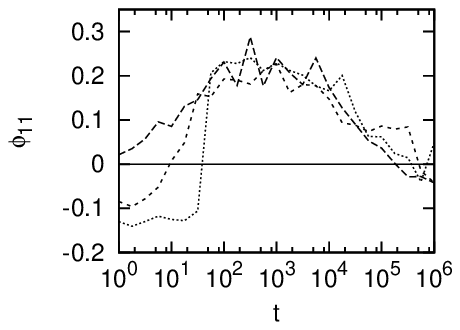} & \includegraphics[width=7cm]{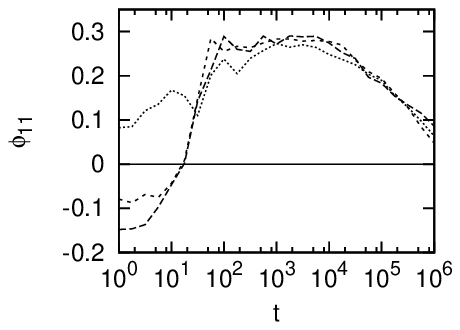} 
\end{tabular}
\caption{Evolution as a function of time of $\phi_{11}$ for $N=100$ particles (left panel) and $N=800$ particles (right panel)
in simulations starting from three different realizations of initial conditions with $R_{0}=0$. An average in a temporal window of width 
$\Delta t=10$ has been used in all cases. Despite the time averaging there are still significant fluctuations which limit
the precision of the determination of the time scale for relaxation. }
\label{fig_h}
\end{center}
\end{figure}

Shown in Fig.~\ref{fig_h} is a plot of the
evolution of $\phi_{11}$ in three different realizations
for $N=100$ and $N=800$ and $R_0=0$, up to $t=10^{6}$.
The quantities are again averaged in the same window as above.
The variance, albeit clearly decreasing with $N$, is in fact
still so significant as to make an accurate determination of
the scaling difficult. Averaging over larger time windows 
the curves become smoother, but such differences persist
if we use a time window which is small compared to the
 time scale of the relaxation itself. In short the intrinsic 
finite $N$ fluctuations from realization to realization  in 
the ($N$ dependent) relaxation time are still so large for
$N$ of this order as to limit significantly the determination of the 
average behaviour from a single realization.

We thus consider a simple ensemble average, over realizations
of the initial conditions. While we could combine time averaging
and such an ensemble average, we choose not to do so as 
this may complicate the interpretation of our result. More precisely,
if we perform a time average, we would need to check carefully
for any possible dependence of our results on the chosen 
averaging window. We will explore below in some detail the relation 
between time averages and ensemble averages over initial conditions.

\begin{figure}[h!]
\begin{center}
\begin{tabular}{cc}
  \includegraphics[width=7.5cm]{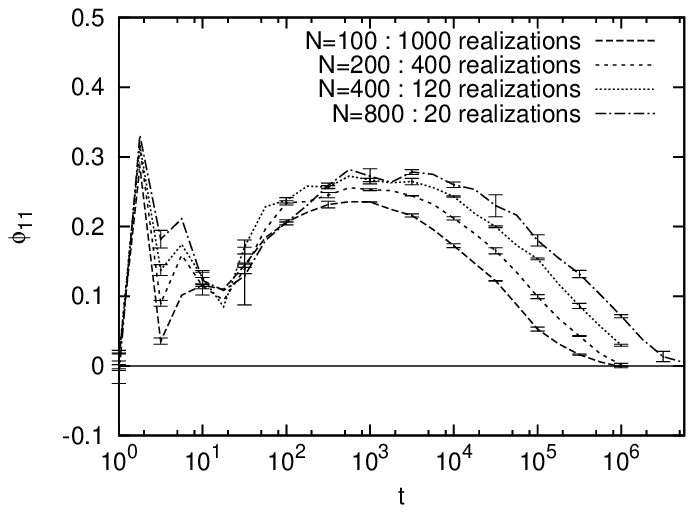} & \includegraphics[width=7.5cm]{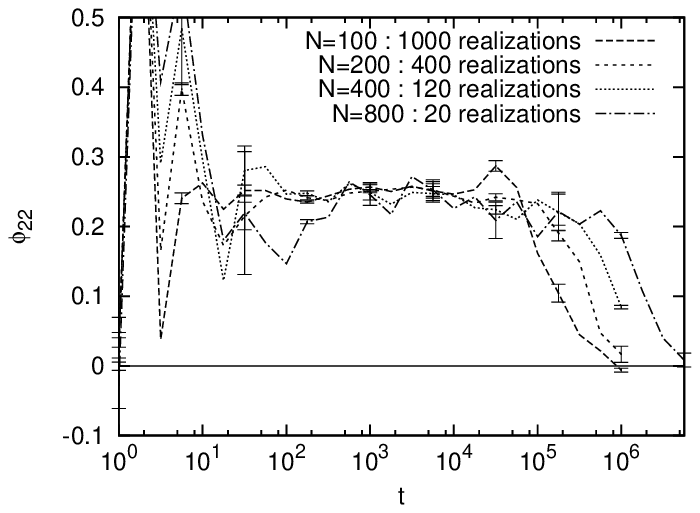} 
\end{tabular}
\caption{Evolution as a function of time $\phi_{11}$ (left panel) and $\phi_{22}$ (right panel),
averaged in all cases over an ensemble of simulations starting from initial conditions
which are realizations of cold uniform initial conditions (i.e.  $R_{0}=0$).
The number of particles $N$ and number of realizations averaged over in each case
is indicated in the panel. The error bars shown are derived (see text) by determining 
the same quantities in two randomly constituted sub-ensembles.}
\label{fig_i}
\end{center}
\end{figure}

Shown in Fig.~\ref{fig_i} are plots of 
$\phi_{11}$ and $\phi_{22}$ averaged over the indicated
number of realizations (and without any time average) 
for each of the indicated
particle numbers, for $R_0=0$.
The error bars in this plot have been 
estimated by dividing randomly the realizations
into two subsamples and recomputing the average
in each of them (i.e. the error bar corresponds
to the difference in the two averages). 

\begin{figure}[h!]
\begin{center}
 \includegraphics[width=9cm]{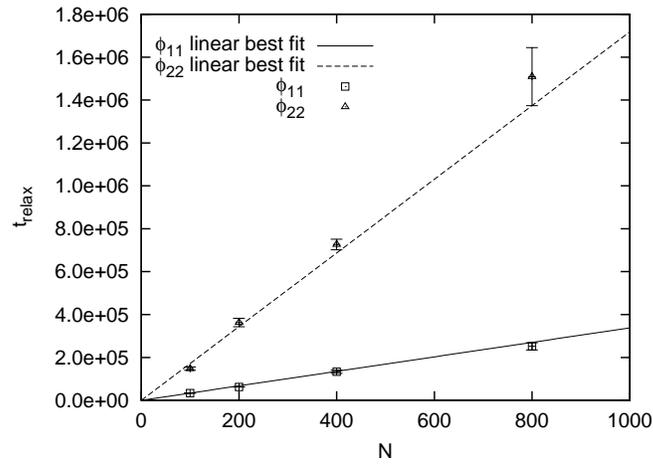}
\caption{Relaxation time as a function of $N$, estimated
as described in the text from the ensemble averaged  evolutions 
of $\phi_{11}$  and $\phi_{22}$ shown in the previous figure. 
Linear best fit lines are also shown. The error bars indicated 
are derived from those in the preceding figure.}
\label{fig_j}
\end{center}
\end{figure}

Using these results we now determine the scaling
with $N$. Shown in Fig.~\ref{fig_j} 
is a plot of  of which  $t_{relax}$,  the characteristic time scale 
for relaxation, as a 
function of $N$ estimated from each of the curves 
for $\phi_{11}$ and $\phi_{22}$. We have determined
the value of  $t_{relax}$ in each case as that at which
the order parameter reaches half its ``plateau'' value 
(i.e. in the QSS), i.e., we estimate the value of
the parameter which corresponds to the approximate
plateau and then determine the time at which
half this value is attained. The error bars correspond to 
those estimated from those given in the previous figure.
Shown also are linear behaviours, which in both cases 
provide a good fit to the results.

\begin{figure}[h!]
\begin{center}
\begin{tabular}{c}
  \includegraphics[width=7.5cm]{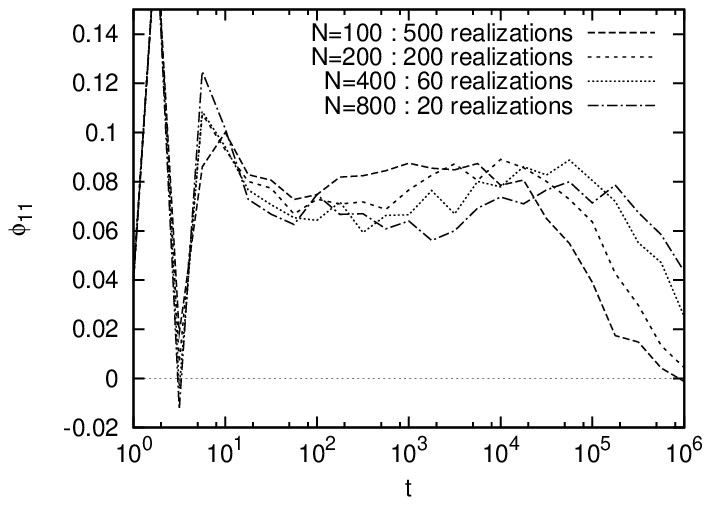} \\ 
\begin{tabular}{cc}
  \includegraphics[width=7.5cm]{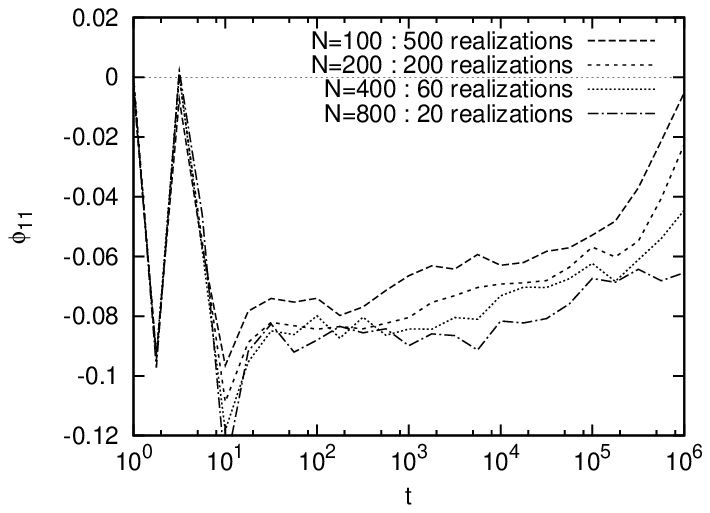} & \includegraphics[width=7.5cm]{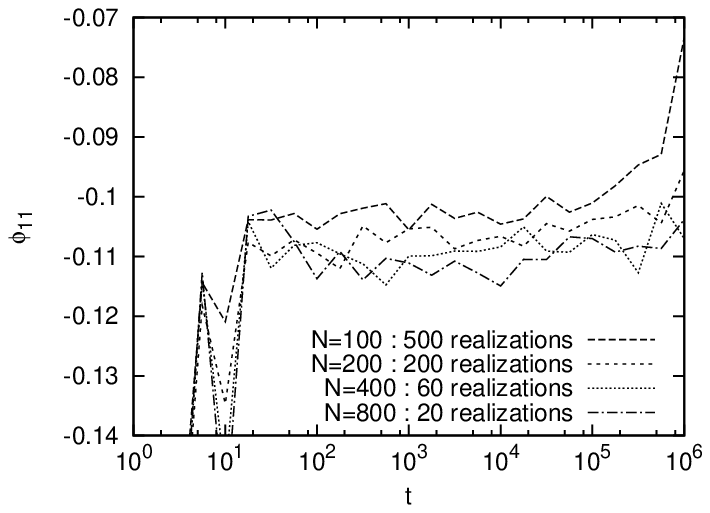}
\end{tabular}
\end{tabular}
\caption{Evolution as a function of time of $\phi_{11}$ averaged over an ensemble of simulations 
started from initial conditions with $R_{0}=0.1$ (top panel),  with $R_{0}=0.5$ (bottom-left panel) and 
with $R_{0}=1$ (bottom-right panel). The number of particles $N$ and realizations in each case is 
indicated in the panels.} 
\label{fig_k}
\end{center}
\end{figure}

\begin{table}[h!]
\caption{Estimated relaxation time $t_{relax}$ for different initial conditions. ``-"indicates that our data does not allow us 
a determination of this time using our chosen criterion.}
\begin{tabular}{ccccc}
\hline\hline 
$N$ & $R_{0}=0$ & $R_{0}=0.1$ & $R_{0}=0.5$ & $R_{0}=1$ \\ 
\hline 
100 \ \ & \ \ 3.4e+04 \ \ & \ \ 8.4e+04 \ \ & \ \ 2.7e+05 \ \ & \ \ - \ \ \\ \vspace{0.1cm}
200 \ \ & 6.5e+04 & 1.7e+05 & 5.3e+05 & - \\ \vspace{0.1cm}
400 \ \ & 1.3e+05 & 6.8e+05 & - & -\\ \vspace{0.1cm}
800 \ \ & 2.5 e+05 & - & - & - \\ 
\hline\hline
\end{tabular}
\label{table-1}
\end{table}

Shown in Fig.~\ref{fig_k} are the ensemble averaged evolution of
$\phi_{11}$ for the three other initial  conditions, for the
same four values of $N$.  The determinations of the relaxation times, for 
each case where this is possible by the same method as used above, 
are shown  in Table \ref{table-1}.  As there are so few points we have 
not performed the same fitting procedure (with estimated error bars)
as above, but it is clear that the given values are consistent with
a scaling of the relaxation time linear in $N$. In the case $R=1$, however, 
we cannot deduce any reliable estimate of the scaling with $N$,
as we can just see the onset of the relaxation for $N=100$
but not in the other cases.

This last curve and the data in Table \ref{table-1} allow us to see more 
quantitatively the dependence of the relaxation time on the initial
value of $R_0$ also. At fixed $N$ we see that, between $R_0=0$
and $R_0=0.5$ the estimated relaxation time increases by a 
factor of about eight. These considerable differences translate
into a very different appearance to the curves: in the case
of $R_0=0$ the ``QSS plateau" is much less visible as there
is only a very small separation between the time scales for the 
establishment of the QSS ($\sim 10^2$) and the onset of 
relaxation.

The exact definition taken here for the relaxation time
is somewhat arbitrary --- we could equally consider
the time at which  $\phi_{11}$ deviates by $10\%$
from its plateau value, or, say,  reaches $10 \%$ of
this value.  Because the relaxation is very slow 
--- to show the evolution of $\phi_{11}$ we must 
plot it as a function of the logarithm of time ---
such definitions would give enormously different 
results for the estimated time (differing by two
to three orders of magnitude). Equally we see 
from  Fig.~\ref{fig_j} that if we use $\phi_{22}$
rather than  $\phi_{11}$, employing the same
criterion we obtain times differing by an
order of magnitude. That this factor indeed changes only 
the overall normalization  of the times, and not their 
scaling with $N$, is evident from the fact that, as can 
be seen by eye, the curves in the decay phase 
can be superimposed on one another well by a 
translation parallel to the time axis. 

\begin{figure}[h!]
\begin{center}
  \includegraphics[width=9cm]{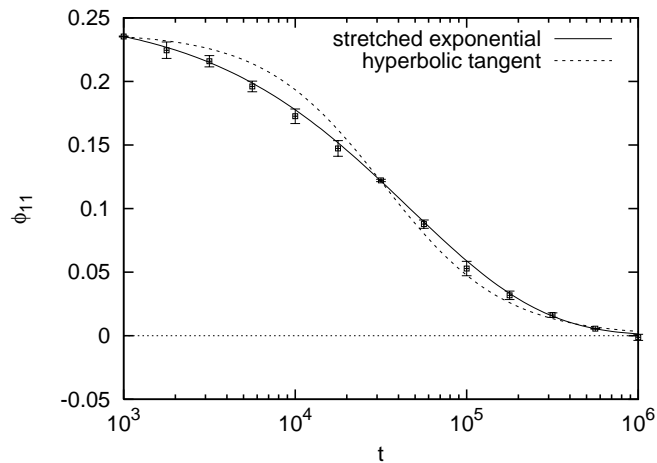} 
\caption{ Evolution as a function of time of $\phi_{11}$, for $t > 10^3$, 
calculated over an ensemble of one thousand realizations of initial
conditions with $R_{0}=0$ sampled by $N=100$ particles.
A best-fit to a hyperbolic tangent [see Eq.~(\ref{fit-hyperbolic})] 
is shown as a dashed line, while the solid line is that  for a stretched 
exponential form [see Eq.~(\ref{fit-stretched})]. The error bars 
are the same as those in Fig.~\ref{fig_i}.} \label{fig_u}
\end{center}
\end{figure}

It is interesting to see if a simple functional
behaviour can be fit to the decay of  the order
parameters. Shown in Fig.~\ref{fig_u}
are best fits to two simple functions for the case
of initial conditions $R_0$ and $N=100$,
for which we have the best statistics. We
have restricted to the range 
$t>10^{3}$ to cut out the initial (violent)
relaxation phase. One employs a hyperbolic 
tangent given by
\be
\frac{\phi_{QSS}}{2} \big\{1-\textrm{tanh}[\alpha_h (\textrm{log}\, t-\textrm{log}\, t_{relax})]\big\}\label{fit-hyperbolic}
\ee
in which, therefore, $t_{relax}$ corresponds to the time estimated above.
The best fit values of the parameters are $\phi_{QSS}=0.24$, 
$t_{relax}=10^{4.5}$ and $\alpha_h=1.4$. The other is a stretched 
exponential form:
\be
\phi_{QSS}  \exp\big\{ -[ \frac{t}{t'_{relax}}]^{\alpha_s}\big\}
\label{fit-stretched}
\ee
and gives the best-fit values $\phi_{QSS}=0.26$
$t'_{relax}= 10^{4.7}$ and $\alpha_s=0.56$. The second function is clearly a significantly 
better fit. We note that the former function has been shown in 
\cite{yamaguchi_etal_04}  to give a good fit to
the temporal evolution of the magnetisation
during relaxation in the HMF model. Stretched exponential
relaxation, on the other hand, is observed in a range of  physical systems, and 
notably in the relaxation of structural and spin glasses (see, e.g.,  \cite{almeida_etal_2001}).

We draw attention to one important feature of these results
which introduces a systematic uncertainty into them, which could
only be reduced by doing significantly larger simulations:
in principle the intermediate QSS is independent of the number 
of particles $N$, i.e., we are estimating the $N$ dependence
of the relaxation time of a state which is, up to fluctuations,
$N$-independent; in practice it is clear in our
data that there is some residual $N$ dependence in the
QSS at the $N$ we are simulating --- the ``plateau"  in  
the curves of the time evolution of our order parameters 
do not exactly coincide. As we have seen that there is
clearly a significant dependence of the lifetime on
$R_0$, which we interpret to be one on the intermediate
QSS rather than the initial condition itself, it is possible that
the $N$ dependence we measure at fixed $R_0$ is due
to, or partially due to, this residual $N$ dependence of
the QSS. We believe, however, that such an effect, if
present, is probably negligible:  the differences in
the QSS  ``plateau" at given $R_0$ for different $N$ 
are very small compared to the differences between 
the QSS over the range of $R_0$, and further, for
the larger $N$, the QSS do appear to converge.
This is even the case for $R_0=0$, where the 
$N$ dependence in the ``plateau" is most evident. 
In this case, as we mentioned above when we discussed 
our initial conditions, an intrinsic  $N$ dependence of
the QSS might be anticipated: as $N \rightarrow \infty$
the evolution becomes singular at $t=1$, and
the evolution at finite $N$ is regulated by the
fluctuations about a uniform distribution 
which are $N$ dependent\footnote{For the analogous
3D problem --- evolution from cold uniform initial 
conditions --- the precise $N$ dependence of the
virialized QSS state has been determined numerically
in  \cite{ejection_mjbmfsl}.}.  That such intrinsic 
$N$ dependence is weak, if present at all, is also
indicated by the absence of visible $N$ dependence
of $\phi_{22}$ in Fig.~\ref{fig_i}.

\subsection{Relaxation and fluctuations in the QSS}

Analytically the relaxation towards equilibrium of systems
with long range interactions may be described by kinetic 
equations, derived for example from the BBGKY hierarchy. 
In practice these equations are intractable, and despite
many attempts to develop appropriate approximation schemes
which might make them tractable, there are really no solid
results which allow us rigorously to model analytically
the detailed phenomenology of relaxation observed in numerical
simulations, and determine for example the observed $N$
dependence of the relaxation time.

Inspection of our results for the temporal evolution
of the parameters $\phi_{11}$ and $\phi_{22}$ lead to
one simple observation: the relaxation time appears to be
correlated with the amplitude of the fluctuations about
the relevant QSS, i.e., the smaller the fluctuations in
the QSS, the longer is its lifetime. While this is somewhat trivial 
when we consider a given initial condition (i.e. $R_0$) at fixed $N$ ---
in postulating that there is a QSS we mean that the
fluctuations about it are $N$ dependent (and decaying
with $N$) --- it is not evident that this should be so 
for the different $R_0$ at fixed $N$. Theoretically
such a correlation might not be surprising --- in kinetic
theory approaches  the leading corrections to the 
collisionless (Vlasov) limit are, in perturbative approaches,
sourced by fluctuations about the QSS (see, e.g., contributions 
of P.H. Chavanis, and of  F. Bouchet and J. Barr\'e in \cite{Assisi}.).

\begin{figure}[h!]
\begin{center}
  \includegraphics[width=9cm]{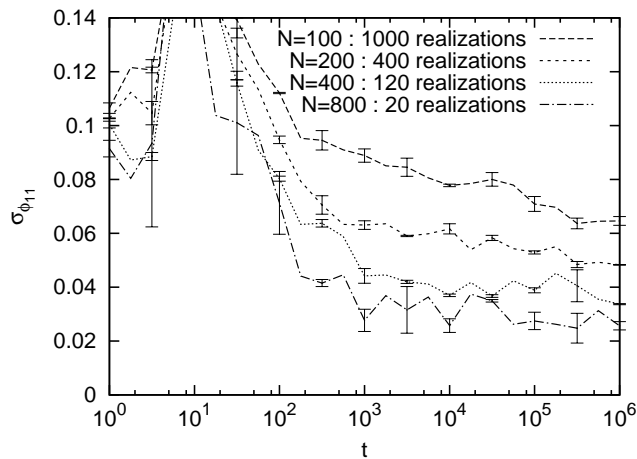}
\caption{Evolution as a function of time of $\sigma_{\phi_{11}}$, the standard deviation 
of $\phi_{11}$,  calculated in a set of simulations starting from independent realizations of initial 
conditions with $R_{0}=0$. The different curves correspond, as indicated,
to different values of $N$ and numbers of realizations. The error bars are 
derived by randomly dividing the set of simulations in each case into
two subsets. The  amplitude of $\sigma_{\phi_{11}}$ at any time
clearly decreases as $N$ increases. }
\label{fig_n}
\end{center}
\end{figure}

\begin{figure}[h!]
\begin{center}
\begin{tabular}{cc}
  \includegraphics[width=7.5cm]{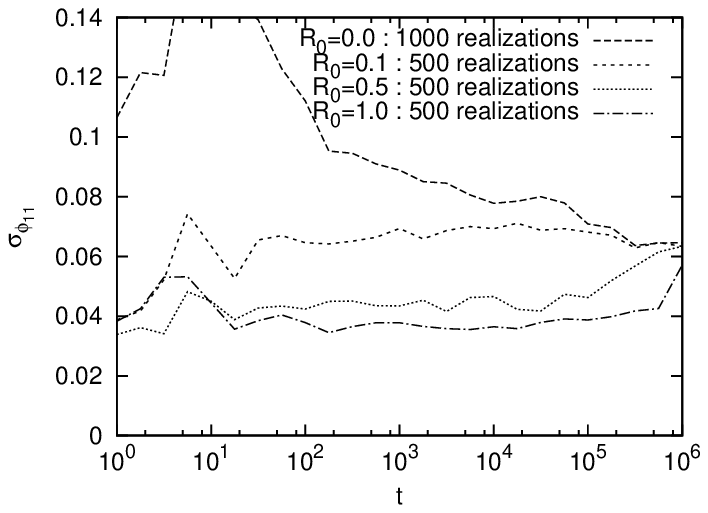} & \includegraphics[width=7.5cm]{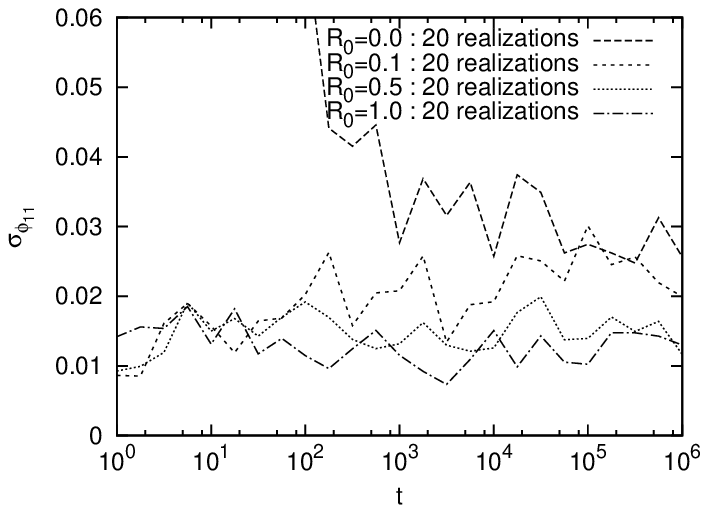}
\end{tabular}
\caption{Evolution as a function of time of $\sigma_{\phi_{11}}$, the standard deviation of $\phi_{11}$, 
calculated in a set of simulations starting from independent realizations of initial 
conditions with the indicated values of $R_0$, sampled with $N=100$ particles
(left panel) and   $N=800$ particles (right panel). The number of realizations 
used in each case is indicated. We observe that in each panel the amplitude of $\sigma_{11}$ 
in the QSS phase is apparently correlated with the duration of this phase, larger 
$\sigma_{11}$ being associated with a shorter relaxation time. In the upper plot the
relaxation to thermal equilibrium is reflected in the convergence of $\sigma_{11}$ 
for different $R_0$ at later times.}
\label{fig_o}
\end{center}
\end{figure}

Such a trend can be seen a little in Fig.~\ref {fig_g}, although 
in this case it is greatly obscured by the time averaging
(i.e. it is much clearer if one plots a single realization
in each case, which we have not done here). 
It is shown clearly to be present by the results
in Figs.~\ref{fig_n} and \ref{fig_o}. The first shows 
the standard deviation, $\sigma_{\phi_{11}}$,
of $\phi_{11}$ as a function of time, estimated 
in the indicated number of realizations of initial
conditions $R_0=0$, for each of  the different values 
of $N$ indicated. The error bars in the plot correspond
to the spread in  $\sigma_{\phi_{11}}$ when it is 
estimated in two sub-ensembles defined by randomly
dividing the realizations into two.  As remarked above
the fact that $\sigma_{\phi_{11}}$ decreases with
$N$ --- and thus, given that the lifetimes of the states
have been observed to increase with $N$, that
there is a correlation of the lifetime with their 
amplitude --- is not surprising: it simply means
that the fluctuations about the QSS are, predominantly,
due to finite $N$ effects which will vanish as 
$N \rightarrow \infty$. We note that the amplitude
of $\sigma_{\phi_{11}}$ in the approximate 
``plateau" region --- corresponding to the QSS --
scales as $1/\sqrt{N}$, i.e., as they would if the
fluctuations of $\phi_{11}$  is the sum of $N$ 
uncorrelated contributions from the $N$ 
particles. 

Shown in Fig.~\ref{fig_o} is the same quantity but now for the different values of
 $R_0$, at two different fixed $N$ ($N=100$ and $N=800$). In both cases
 we see clearly (except perhaps for the lowest curve in the lower panel,
 which is noisier due to the much smaller number of realizations) that
 the amplitude of fluctuations decreases as $R_0$ increases, i.e., that
 the amplitude is (inversely) correlated with the lifetimes we have
 observed for these states.  
  
We note that these figures giving the behaviour of the variance of
our macroscopic parameters also contain alot of other useful information
beyond the correlation we have just observed. Indeed these curves
themselves show very clearly the different time-scales in the 
dynamics: the first period of ``violent relaxation" is clearly
identifiable by a very large variance, which decays on a time
scale of order several tens of dynamical times; this is followed
by an approximately stable value depending on the QSS
(Fig.~\ref{fig_o}), which then evolves on a much  longer time
scale, dependent on and increasing with $N$,  towards a value 
which is independent of the initial state (i.e.  thermal equilibrium). 

These results also allow us to conclude more about the meaning of
our quantitative results for the scaling of the relaxation time, which
have been calculated using the ensemble average:  the systematic decrease 
with $N$ of the variance of $\phi_{11}$ (and, we have verified, of $\phi_{22}$)
implies that such an ensemble average, for sufficiently large $N$, can 
indeed be interpreted consistently to give the macroscopic behaviour 
of a single realization in the ensemble. Thus, as we have been
implicitly assuming, we can indeed take our determined relaxation 
times to represent those of single realizations, at sufficiently large $N$.

\begin{figure}[h!]
\begin{center}
  \includegraphics[width=9cm]{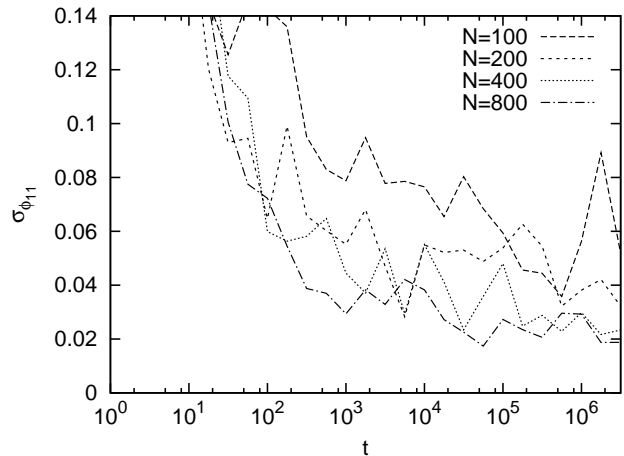}
\caption{Evolution as a function of time of $\sigma_{\phi_{11}}$, the standard deviation of $\phi_{11}$, 
for the same cold initial condition and values of $N$ as in Fig.~\ref{fig_n}, but calculated now
by sampling the value of $\phi_{11}$ at one hundred equally spaced intervals in a temporal
window of width $\Delta t=10$ centred at the indicated time.  We see that values are 
comparable to those in Fig.~\ref{fig_n}, and show the same trend with $N$.}
\label{fig_p}
\end{center}
\end{figure}

\begin{figure}[h!]
\begin{center}
\begin{tabular}{cc}
  \includegraphics[width=7.5cm]{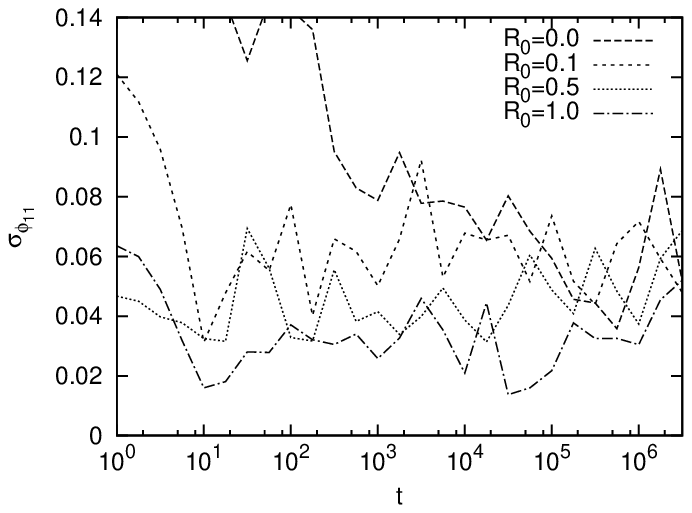} & \includegraphics[width=7.5cm]{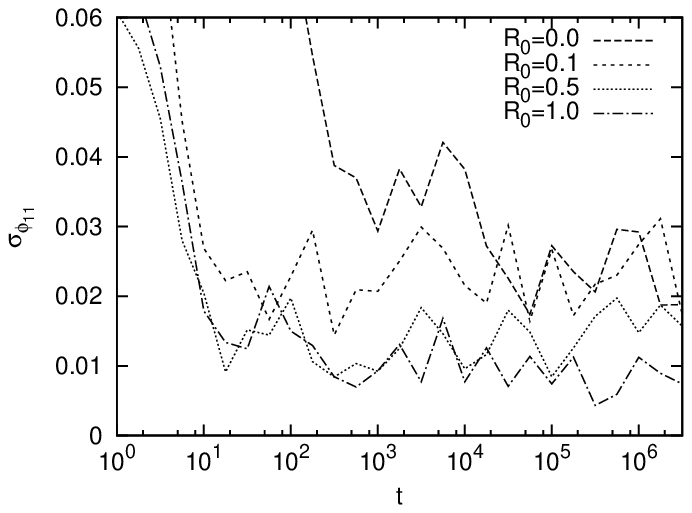}
\end{tabular}
\caption{Evolution as a function of time  of the $\sigma_{\phi_{11}}$, the standard deviation of $\phi_{11}$, 
for the same cases as Fig.~\ref{fig_o},  but calculated now
by sampling the value of $\phi_{11}$ at one hundred equally spaced intervals in a temporal
window of width $\Delta t=10$ centred at the indicated time. The results are very consistent
with those in Fig.~\ref{fig_o}.}
\label{fig_q}
\end{center}
\end{figure}

It is interesting to go a little further and consider what the relation is,
at finite (but large) $N$, between the fluctuations in the ensemble 
average and the temporal fluctuations in a single realization.
Indeed {\it if}, as we have postulated above,  there is a real correlation 
between the amplitude of the fluctuations measured in the ensemble average 
and the lifetime of the corresponding QSS,  it must be that these
fluctuations measured in the ensemble bear some close relation to the 
temporal fluctuations in the same parameters in  a single realization. 
That this is the case, to a good first approximation, can be seen from  Figs.~\ref{fig_p} 
and \ref{fig_q}, which show exactly the same
quantities as in the previous two figures, but that
the standard deviations are calculated in
one hundred time slices equally spread over
a time window of width $\Delta t=10$ centred
on the indicated point. We see the same quantitative
behaviours as in the previous plots, and even,
in particular for $N=800$, quite  good qualitative
agreement.  

\begin{figure}[h!]
\begin{center}
\begin{tabular}{cc}
  \includegraphics[width=7.5cm]{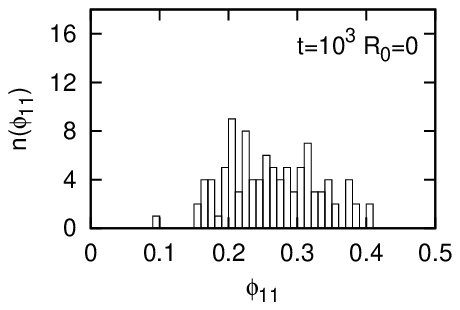} & \includegraphics[width=7.5cm]{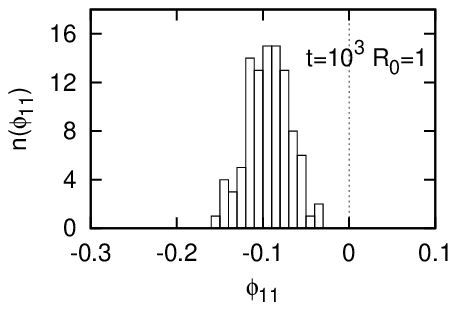} \\
  \includegraphics[width=7.5cm]{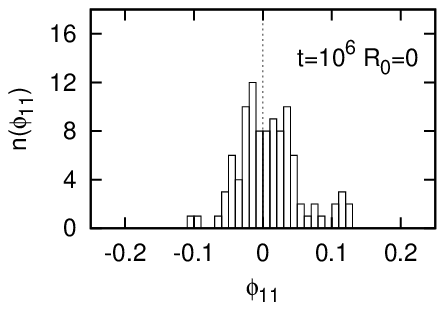} & \includegraphics[width=7.5cm]{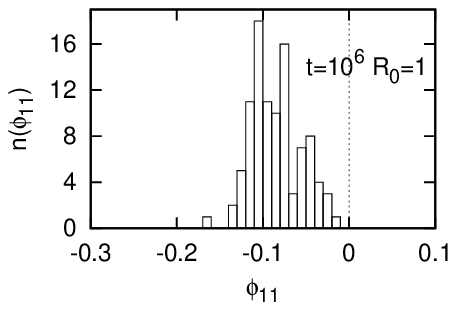}
\end{tabular}
\caption{Histograms of the values of $\phi_{11}$ measured at the indicated
time in one hundred simulations started from independent realizations 
of initial conditions with the indicated values of $R_0$, for $N=200$ particles.
The dashed line indicates $\phi_{11}=0$, the value at thermal equilibrium.
For $R_0=0$ the relaxation of the system from the QSS to thermal equilibrium
is clearly visible, with both the central value and shape of the distribution evolving.
For $R_0=1$ we observe, in line with the results above, that the system is still
in a QSS but that the distribution has started to shift slightly towards the
equilibrium one.}
\label{fig_r}
\end{center}
\end{figure}

\begin{figure}[h!]
\begin{center}
\begin{tabular}{cc}
  \includegraphics[width=7.5cm]{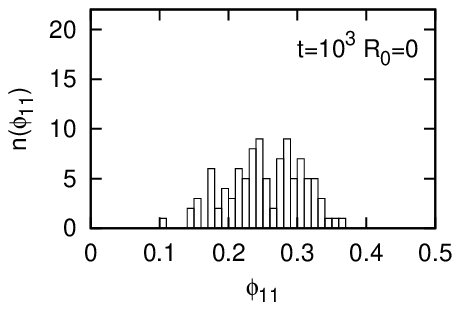} & \includegraphics[width=7.5cm]{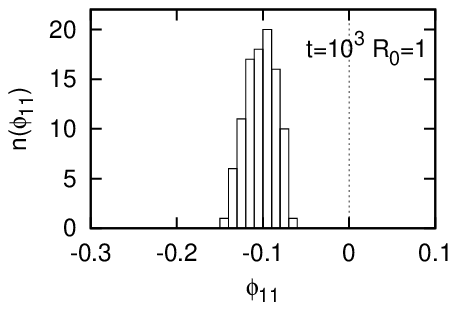} \\
  \includegraphics[width=7.5cm]{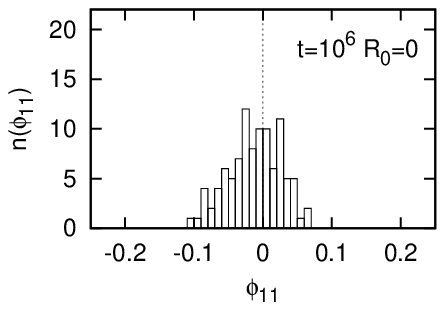} & \includegraphics[width=7.5cm]{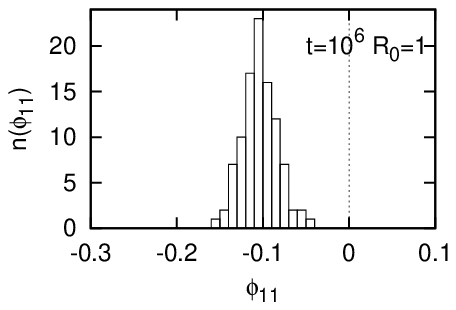}
\end{tabular}
\caption{Histograms of the values of $\phi_{11}$ measured at one hundred equally spaced 
times in a temporal window of width $\Delta t=10$ centred on the time indicated in each panel.
The simulations are from the same cases as in Fig.~\ref{fig_r}. We observe a qualitative
agreement between the amplitudes and shapes of those in Fig.~\ref{fig_r}.}
\label{fig_s}
\end{center}
\end{figure}

To allow further more detailed comparison, in 
Fig.~\ref{fig_r}  and Fig.~\ref{fig_s}  are shown,
for $R_0=0$ (left panels) and $R_0=1$ (right panels),
and $N=200$ in both cases, the histogram of the values of $\phi_{11}$, 
at the indicated times measured, in Fig.~\ref{fig_r},  in one hundred simulations from 
realizations of the same initial conditions,  and, in Fig.~\ref{fig_r}, in one hundred 
snapshots in a window of width  $\Delta t=10$  in a single simulation from
one realization of the same initial conditions. Comparing the fours panels
in the two figures one by one, we see that,  although the fluctuations in each case 
are clearly not sampled from an identical distribution, the agreement 
is strikingly good: not only, as expected from what we have already
seen above, do the averages and variances agree well
in each case, but the general shape of the 
histograms, which is quite different in each QSS,  
resemble one another strongly.  The results are also clearly 
in line with the conclusion drawn above for what concerns
the relaxation to thermal equilibrium: at $t=10^6$ we see
that the cold initial conditions have relaxed to a distribution
centered on the value in thermal equilibrium, while for
the case $R_0=1$ the system is still in a QSS but has
evolved very slightly towards equilibrium.

It would be interesting to  develop this study with
greater statistics, varying the width of the time window to see how
good agreement can be obtained, but we will not do so
here.  Such a study is related to the fundamental 
(and so far unanswered) question as to whether 
the properties of QSS may be determined by  
averaging over an appropriate (non-equilibrium)
ensemble, determined by the initial conditions. 
The theory of violent relaxation formulated
by Lynden-Bell, for example, postulates an answer to
this question \cite{lyndenbell}: the appropriate ensemble is that of
all configurations corresponding to a phase
space distribution function permitted by 
the (collisionless) Vlasov dynamics. If 
this theory were correct (which is not the
case for this system \cite{yamaguchi_2008})
we should perform our ensemble 
average over such configurations rather than the more 
restricted one we have considered. 

\section{Conclusions and discussion}

\subsection{Summary}

Our primary aim in this paper has been to establish and characterize 
more fully than in the previous literature the relaxation to thermodynamic equilibrium 
of one of the simplest toy models for long-range interacting systems:
equal mass self-gravitating  particles in one dimension (or infinite sheets in three dimensions).
Compared to the much studied HMF model, notably, the basic properties of this model 
have remained somewhat unclear, and indeed marked by some controversy in the 
literature. The novelty of our work compared to previous studies is not 
just that we do more and larger simulations from a broader range of
initial conditions, but that we have identified a tool which is very useful 
to characterize the evolution of the system: the measurement of
appropriately normalized moments of the distribution function which
characterize the ``entanglement" of the one particle distribution function
in configuration and velocity space.  This is particularly appropriate as a 
measure simply because the thermal equilibrium has the property that 
such entanglement is absent while, we have shown, in {\it any other} 
stationary solution of the Vlasov-Poisson equations such 
entanglement is present. We note that this result, which we showed
to be valid for any interaction in one dimension (but, as noted,
excluding periodic systems like the HMF),  can be generalized
easily to three dimensions if we restrict to stationary solutions which 
have radial symmetry in space and velocity. This suggests that
these ``order parameters" may also be useful indicators of
relaxation in much more general, and perhaps, as we discuss
below,  even  useful tools for understanding the mechanisms of such 
relaxation.

With the aid of these macroscopic measures, we have shown in our 
numerical study,  of a range of  simple ``waterbag" and cold initial 
conditions,  that the system manifests the behaviour thought to be 
generic in long-range systems: there are essentially two phases
in the evolution with two completely different time-scales. An
initially non-stationary state evolves first,  on timescales characterized
by the ``dynamical time"  $t_{\rm dyn}$ (roughly the crossing time of a particle 
in the mean-field due to the others),  to a QSS, an out of equilibrium
state, which then evolves on a much longer time scale, dependent
on the number of particles,  to thermal equilibrium. In other words
it is reasonable to suppose that the system is ergodic (and mixing) on 
these very long time scales, but not so on the shorter time scales. 
Further we can identify clearly that the QSS resulting from different 
initial conditions (i.e. different values of $R_0$)  are very different 
macroscopically, characterized by very different phase space 
entanglement. 

Focussing on the the $N$ dependence of the relaxation, averaging 
over a very large number of realizations to average out the fluctuations, 
we have concluded that the characteristic time scale for relaxation
behaves, to a very good approximation, as 
\be
t_{\rm relax}  \sim  f_{QSS} \, N \, t_{\rm dyn}
\label{relaxation-scaling}
\ee
where $f_{QSS}$ is a numerical factor which depends on the
initial condition, which we have denoted in this way as we 
expect that this dependence is not strictly on the initial
condition but on the QSS which results from it.  We have
seen that this prefactor increases as $R_0$ does, by about 
a factor of ten between $R_0=0$ and $R_0=0.5$, and 
approximately a further factor of ten  for $R_0=1$.
We have noted that the overall normalization of
$f_{QSS}$ is rather arbitrary, as it depends greatly
on the exact criterion used to define the relaxation 
time-scale.  Given that the evolution
towards zero of $\phi_{11}$, which is what we have 
used to determine this time scale, is in fact well
fit by the simple functional behaviour as a 
function of the time on a logarithmic scale, the 
normalisation of $f_{QSS}$ can differ by two orders 
of magnitude  by a trivial change in its definition. 
More specifically we have seen, that in the 
case where we have accumulated the greatest
statistics allowing us to constrain the temporal
evolution, a very good fit to our order parameter
 $\phi_{11}$ is obtained to a stretched exponential form. 

Although the relaxation of this system, and in general
long-range interacting systems, is not well understood,
we can say that this finding of a linear scaling --- besides
the fact that it is, as we will discuss below,  in line with 
less complete previous analyses --- is not a surprising result: 
such a scaling can be anticipated both on the basis of simple 
naive estimates of the effects of collisionality, as well from 
general considerations based on kinetic theory. 

A simple ``derivation" of this scaling, along the lines
of those applied originally by Chandrasekhar (see \cite{chandra43}
or \cite{binney}) to obtain such an estimate for 3D
self-gravitating systems, may be given as follows. Relaxation 
is in principle due to the discretisation, in $N$ particles, of a 
continuous mass distribution. Let us suppose that this 
latter field density varies spatially on a scale, $\ell$.  The typical 
fractional  change in the velocity $v$  of a test particle due to its 
interaction with one (localized) particle, rather than the continuous mass 
distribution,  can be estimated as  $\sim g \ell/m v^2$. As it crosses 
the system (in a time  $\sim t_{dyn}$) such a particle will interact 
with all $N$  particles. Assuming the contribution from each particle 
adds incoherently, one estimates
\be
\frac{\delta v^2}{v^2} \sim N \left(\frac{gl}{mv^2} \right)^2
\ee
for the normalized variance of the velocity in $t_{dyn}$.
Scaling with $N$ at fixed mass and energy (and fixed $\ell$)
requires $g/m \sim 1/N$, and therefore  $\frac{\delta v^2}{v^2} \sim N^{-1}$.
It follows that the relaxation time scales linearly with $N$ in units
of the $t_{dyn}$. A slightly different argument leading to the same 
result may be found in \cite{tsuchiya+gouda+konishi_1996},
and a more developed analysis in \cite{miller_1996}.
In the framework of approaches based on kinetic theory, a linear 
scaling is obtained as collisional terms arise as the leading 
corrections in a formal expansion in $1/N$ which gives the 
collisionless (Vlasov) limit at leading order (see, e.g. 
\cite{balescu,chavanis_KTb_2006, chavanis_kEqns_2010}).

This scaling linear in $N$ is to be contrasted with the case of the scaling 
observed for the life-time in QSS in the HMF, proportional to $N^{1.7}$.
While the exponent found is not understood,  the fact that it is larger 
than unity is consistent with these considerations as this result applies 
for  spatially homogeneous QSS (which are possible in the HMF
because of its periodicity) for which it has been shown that the leading 
collisional term vanishes (see, e.g., contributions of P.H. Chavanis,
and of  F. Bouchet and J. Barr\'e in \cite{Assisi}).

We note that our study suggests also that the ``order parameters" we have 
defined and studied may be relevant quantities for understanding
relaxation in this and other long-range systems. Indeed in all cases we 
have  observed that, at sufficiently large $N$, these parameters start from a 
non-zero value in the initial QSS and evolve monotonically towards 
zero, i.e., the relaxation of the QSS can apparently be described as 
a  process of progressive ``disentanglement'' of the one particle 
phase space density. In this respect the very different, much less
efficient, relaxation observed in the HMF might be interpreted as 
a result of the absence of such entanglement in spatially
uniform QSS.  Further, in the case where we have enough statistics
to provide a precise fit to the evolution of the
parameter $\phi_{11}$, we found that it is well fit by a simple stretched 
exponential form.  It would be interesting to see in further study
whether this fit is more than an approximate numerical
fit for the case we have studied, and, if so, whether the
exponent characterizing it is the same or not. As we have
remarked such a functional form has been observed in other slowly 
relaxing (e.g. glassy) systems and theoretical tools derived in 
this  context to understand relaxation may be relevant.
In \cite{almeida_etal_2001}, for example, this behaviour is
linked to the existence of a fractal structure in a bounded
accessible region of phase space. 

\subsection{Comparison with previous literature}

Let us now turn finally to compare our findings in greater
detail with those in the previous literature. 

An early numerical study by Hohl and Broaddus 
\cite{hohl+broaddus_1967} which concluded a relaxation time 
proportional to $N^2 t_{dyn}$  was found to be incorrect by 
two groups, who studied the problem in greater detail (and with greater numbers
of particles). However, these groups found conflicting 
results:  Miller et al.  found no evidence for  relaxation at all to thermal 
equiibrium in their simulations \cite{wright+miller+stein_1982}, while 
Luwel et al. \cite{luwel+severne+rousseeuw_1983} found 
relaxation on a time scales even shorter than $N t_{dyn}$.  
Further study (see \cite{reidl+miller_1987, reidl+miller_1991},
which also contain a detailed synthesis of the previous
literature) by Miller et al. concluded that the discrepancy
was related to the very specific initial condition studied by the
other group. Studying this case in detail they found that 
it indeed appears to thermalize very rapidly, but some
further, but not completely conclusive analysis of the
evolution at longer times, suggested that this thermalization
was not complete. 

\begin{figure}[h!]
\begin{center}
  \includegraphics[width=9cm]{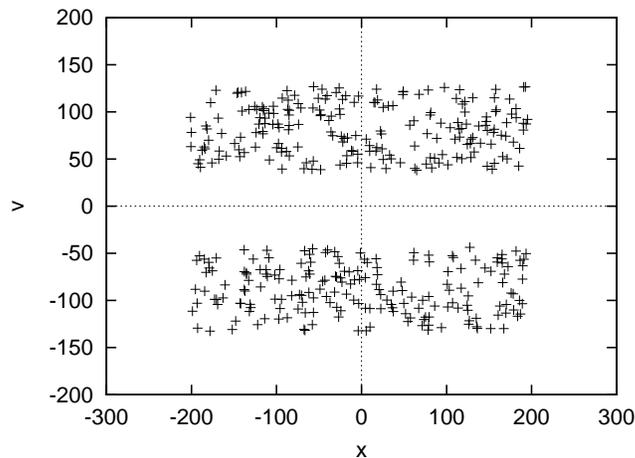}
\caption{A ``counterstreamed'' waterbag initial condition in phase space
with $R_{0}=0.3$, sampled with  $N=400$ particles.}
\label{fig_v}
\end{center}
\end{figure}

\begin{figure}[h!]
\begin{center}
\begin{tabular}{cc}
  \includegraphics[width=7.5cm]{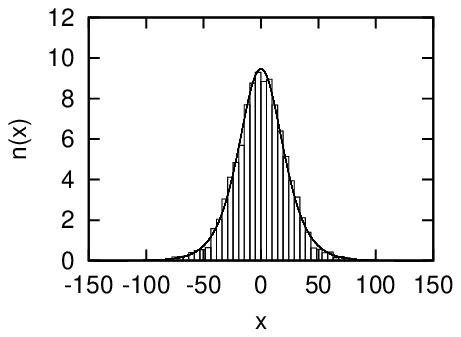} & \includegraphics[width=7.5cm]{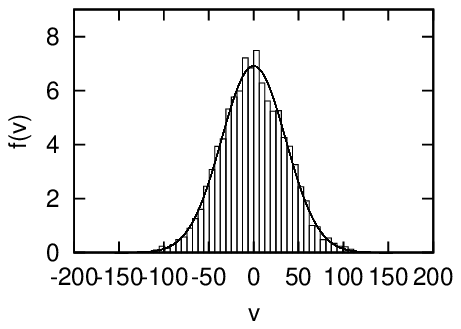}
\end{tabular}
\caption{Density profile (left panel) and velocity distribution (right panel) 
obtained at $t=100$ from a counter-streamed initial condition with $N=100$.
An average over $250$ simulations from independent realizations of the
initial conditions has been performed. The solid lines correspond to the 
values in thermal equilibrium,  Eq.~(\ref{rybicki-eq}).
} 
\label{fig_w}
\end{center}
\end{figure}

\begin{figure}[h!]
\begin{center}
  \includegraphics[width=9cm]{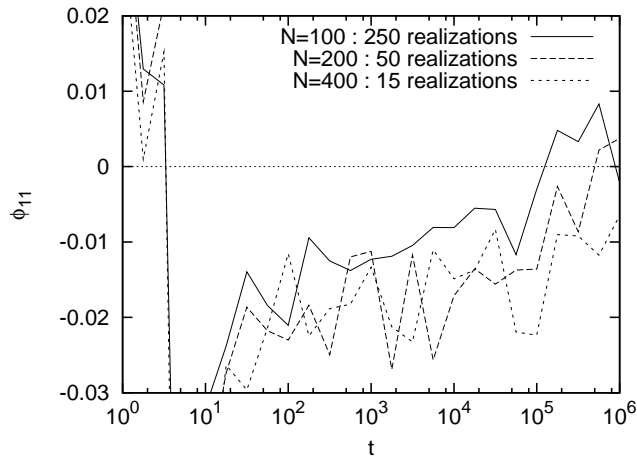}
\caption{Evolution as a function of time of $\phi_{11}$ from
a counterstreamed waterbag initial condition, averaged over the 
number of realizations of the initial conditions and particle
numbers indicated. Despite the indications of the previous
figure, we observe clearly that relaxation to thermal equiibrium
has not taken place at $t=100$.}
\label{fig_x}
\end{center}
\end{figure}

To determine whether these cases are consistent with
our findings --- and see whether our analysis using 
the parameters $\phi_{11}$  and $\phi_{22}$ can 
throw light on these previous findings --- we have 
resimulated the relevant initial conditions. These 
are ``counterstreamed"  waterbag initial conditions, 
an example of which is shown in 
Fig.~\ref{fig_v}. 
We have simulated a range of such initial conditions, 
in particular the cases (one of which is that shown in the figure)
considered by  \cite{luwel+severne+rousseeuw_1983}
and  \cite{reidl+miller_1987}. Shown in Fig. \ref{fig_w} 
are the density profile and velocity distribution 
at $t=10^2$ obtained starting from a realization of 
initial conditions like those shown in 
Fig.~\ref{fig_v}, but with $N=100$.
We see that the profiles indeed agree very
well with the equilibrium ones. In Fig. \ref{fig_x}
is shown the evolution of  $\phi_{11}$ as a 
function of time for the indicated values of
$N$ averaged in each case over the 
number of realizations indicated. 
We observe that, although small and fluctuating, its 
value is clearly on average non-zero, indicating 
that the state, despite the good agreement of the 
density and velocity profiles, is not in fact an 
equilibrium.  Just as in the cases we
studied we see clearly the
relaxation towards equilibrium at longer
times, and indeed that the characteristic time
increases on $N$. Although we haven't 
done the more extensive study required
to determine precisely this $N$ dependence,
the results are quite consistent with 
Eq.~(\ref{relaxation-scaling}) with a value
of $f_{QSS}$ of order that found for
the case $R_0=0$.

This case illustrates  the usefulness of the
parameters $\phi_{11}$ and $\phi_{22}$  
as discriminants of relaxation: indeed 
we have just seen that the single measure
of $\phi_{11}$ is sufficient to discard the
interpretation of Luwel et al. \cite{luwel+severne+rousseeuw_1983}
of their results. This is simply because they are physically
very appropriate indicators, for the
reasons we have explained in introducing
them:   the property they probe --- of 
entanglement of the phase space 
distribution --- is one which must 
evolve significantly during 
relaxation, because the phase
distribution must become separable.
While $\phi_{11}$ and $\phi_{22}$  being 
zero does not {\it imply} thermalization, of
course,  we have not found a single 
QSS, despite exploring a broad range of
initial conditions (considerably more 
extended that those reported here)
in which they are {\it both} zero
(within the uncertainty of fluctuations),
i.e., the only states  we have found in
which  they are  both  zero are states which 
we have concluded, using a range of other 
measures, are indistinguishable  from the
equilbrium state of Rybicki. It is 
not difficult, on the other hand, to find initial 
conditions which lead to a QSS in 
which  $\phi_{11}\approx 0$ {\it or}  
$\phi_{22}\approx 0$. Indeed for the 
waterbag initial conditions we have studied 
both $\phi_{11}$ and $\phi_{22}$
actually change sign as  $R_0$ varies over 
the range we have  considered, and one can
thus find by trial and error the appropriate 
$R_0$ which make them zero individually.

\begin{figure}[h!]
\begin{center}
\begin{tabular}{ccc}
  \includegraphics[width=5cm]{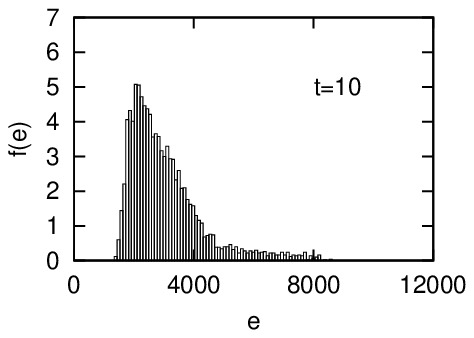} & \includegraphics[width=5cm]{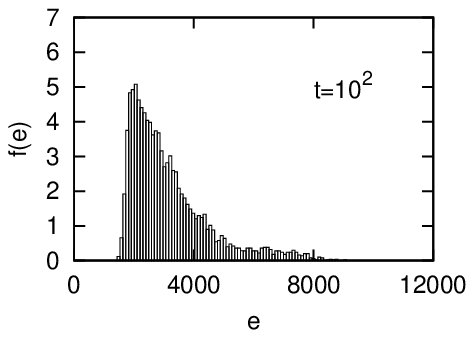} & 
  \includegraphics[width=5cm]{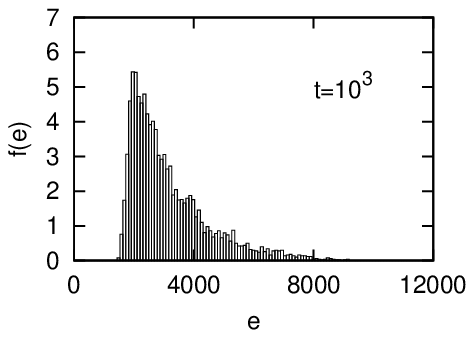} \\ \includegraphics[width=5cm]{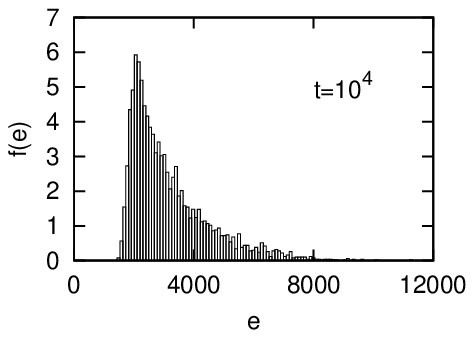} &
  \includegraphics[width=5cm]{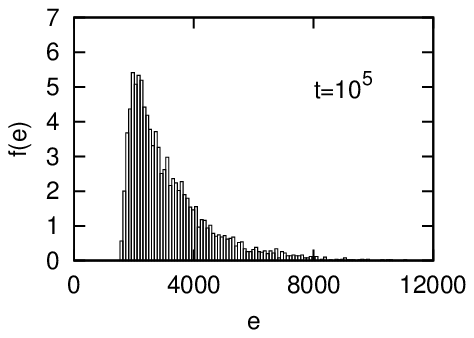} & \includegraphics[width=5cm]{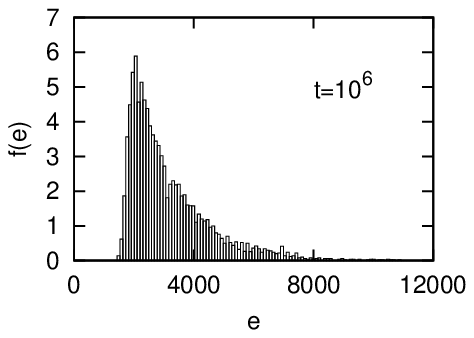} 
\end{tabular}
\caption{Histogram $f(e)$ of individual particle energies $e$ measured at the indicated times 
starting from counter-streamed initial conditions sampled with $N=100$ particles.
The curves are averaged over $250$ realizations. In thermal equilibrium $f(e)$ is 
indistinguishable from the one measured at the latest time shown. The result 
confirms that relaxation in fact occurs on time scales similar to those
observed for the simple waterbag initial conditions.}
\label{fig_y}
\end{center}
\end{figure}
 
Another evident quantity to measure, which we have 
in fact considered systematically but have not reported
in detail, is the distribution $f(e)$ of the individual
particle energy $e$. 
This is in fact generally a better discriminant of 
relaxation than either $n(x)$ and $f(v)$, i.e., we have
found that in quite alot of cases $n(x)$ and $f(v)$
are not easy to distinguish from the equilibrium
profiles, but that $f(e)$ allows one to see more
clearly that one is indeed not in the equilibrium
state.  An example is the counter-streamed case 
just considered above.
In Figs. \ref{fig_y}  are shown, for
$N=100$, the evolution of the ensemble
averaged $f(e)$ at a few different times. We 
have not plotted the equiibrium curve, as it 
is indistinguishable from the measured curve
at the final time shown. One can see clearly
see that, despite the good agreement
of $n(x)$ and $f(v)$ shown in Fig. \ref{fig_w}, the 
system is not in equilibrium at the early times:
$f(e)$ has a clearly visible excess of particles
at high energies compared to that at the 
much later times at which
the evolution of $\phi_{11}$ indicated
relaxation (and $f(e)$ indeed approaches very 
accurately its predicted equilibrium form). While 
such a measure of $f(e)$, averaged over a large ensemble of
realizations,  can, in all the cases for which 
we've studied it, clearly 
discriminate relaxation, the use of just
$\phi_{11}$ (and possibly  $\phi_{22}$) is 
an extremely efficient short-cut to ``diagnose"
relaxation. 

Subsequent to  \cite{reidl+miller_1987}, in the nineties, 
Tsuchiya et al. reported an analysis of larger and
more importantly, longer, simulations in order
to clarify the issue.  A first paper \cite{tsuchiya+gouda+konishi_1996}  
they reported the evolution of a rectangular waterbag initial condition 
corresponding to our case $R_0=1$, and 
reported a detection of relaxation to thermal 
equilibrium. These authors made a distinction
between two time scales of relaxation:  one
of  ``microscopic relaxation'',   the other for
``macroscopic relaxation''. These are identified,
and both found to be proportional to $N$, by considering 
the evolution of the mean standard deviation of 
the particle energies averaged over a time 
window $T$ from their  equipartition value. 
The former  is estimated from the slope at short
time of this function, and the latter from the 
position of ``peaks" which are observed to occur
at much longer times.  While the latter is interpreted
in terms of of macroscopic relaxation in the sense we have
used here, the former is interpreted as a time scale
on which particles sample the energy distribution
but on which there is {\it no} macroscopic evolution.
The justification for these interpretations are 
not clear, and no direct comparison
with the equilibrium distribution derived by
Rybicki, Eq.~ (\ref{rybicki-eq}),  has
been reported which might show their correctness.
Indeed both a subsequent article by the same
authors \cite{tsuchiya+gouda+konishi_1997} and
a study by Yawn and Miller \cite{yawn+miller_erg_1997} 
place in doubt the correctness of the  interpretation in terms 
of relaxation.

\begin{figure}[h!]
\begin{center}
\begin{tabular}{ccc}
  \includegraphics[width=5cm]{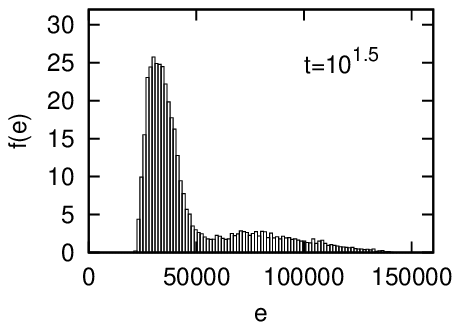} & \includegraphics[width=5cm]{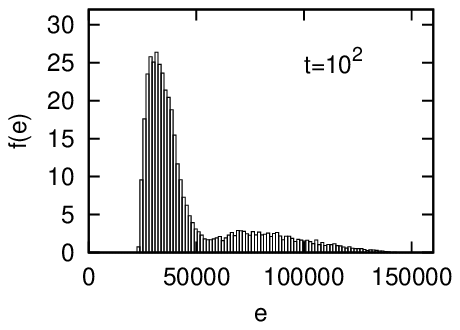} &
  \includegraphics[width=5cm]{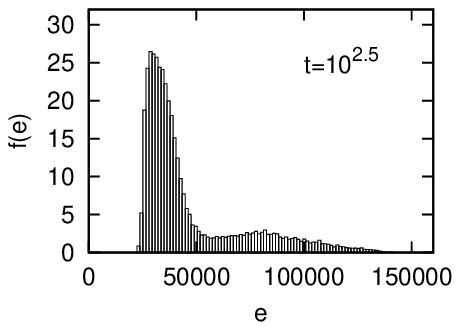} \\ \includegraphics[width=5cm]{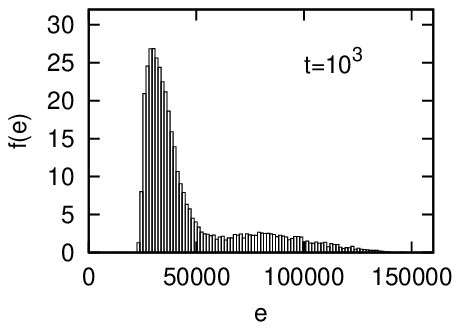} &
  \includegraphics[width=5cm]{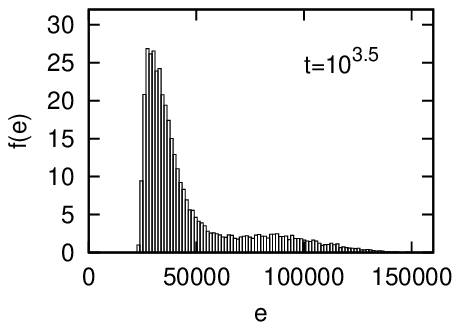} & \includegraphics[width=5cm]{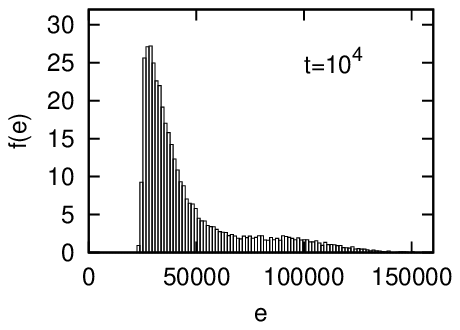} \\
  \includegraphics[width=5cm]{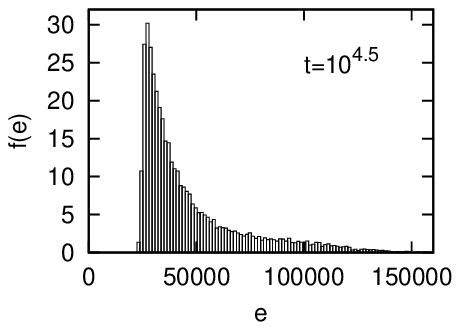} & \includegraphics[width=5cm]{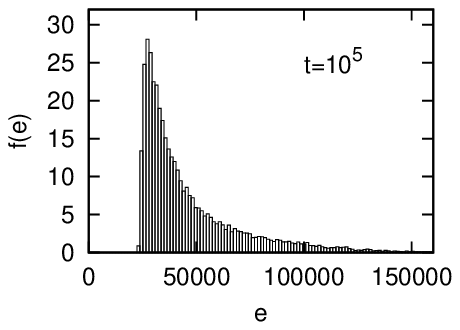} &
  \includegraphics[width=5cm]{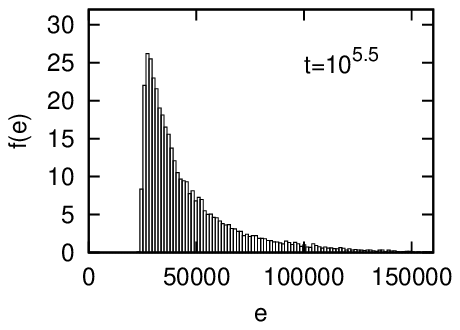} \\ \includegraphics[width=5cm]{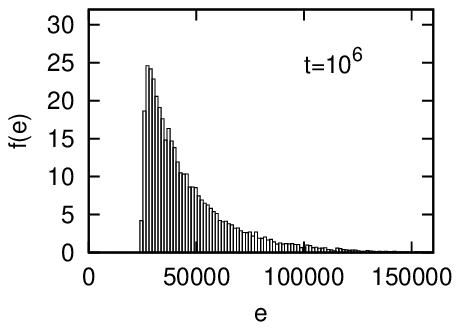} 
\end{tabular}
\caption{Histogram $f(e)$ of individual particle energies $e$ at the indicated time, and
averaged over $60$ simulations from realizations of simple waterbag initial conditions 
with  $R_{0}=0.1$ and $N=400$. The curve at the latest time coincides well 
with that expected in thermal equilibrium. The onset of relaxation is already
visible at a time of order $10^4$,  almost two orders of magnitude smaller 
than the time determined in Table~ \ref{table-1}.} 
\label{fig_z}
\end{center}
\end{figure}

Nevertheless, in light of the results we have given here,
it would be reasonable to infer that the results given
by Tsuchiya et al. in  \cite{tsuchiya+gouda+konishi_1996} are indeed 
correct, at least for what concerns the $N$ dependence
of the relaxation. Further comparison could of course
clarify the relation of the behaviour of their measured
quantities and the macroscopic relaxation as we have 
probed it here (and should be much easier for
the shorter lived, smaller $R_0$, initial conditions
rather than the case $R_0=1$ studied by these
authors). We do not believe, however, that there is any 
clear basis for either an operational or {\it physical} distinction 
between ``microscopic" and ``macroscopic" relaxation as 
described by these authors: as we have discussed 
there is an arbitrariness in the definition of the
relaxation time because of the very slow
nature of this relaxation. As we have noted,
we could easily, for example,  have obtained here estimates 
of the relaxation time differing by several of orders in magnitude in
their prefactor, just like the two different time
scales determined  by Tsuchiya et al.  \cite{tsuchiya+gouda+konishi_1996},
by using slightly different definitions, or choosing
to use a different order parameter.  This point
can be illustrated by considering the evolution
of $f(e)$ for one of the cases we have considered:
shown in Figs. \ref{fig_z} is this quantity
for the case $R=0.1$ and $N=400$, averaged over
$60$ realizations.  While we have associated
(see Table \ref{table-1} above) the time scale 
$7 \times 10^{5}$ to the relaxation in our 
analysis, one can discern by inspection
of these figures significant evolution
(in particular of the initially clear
``core-halo" structure)  in $f(e)$ already 
by $t=10^{3.5}$, i.e., there {\it is} evolution
of the energy distribution on the time
scale of ``microscopic" relaxation (of
order $N t_{dyn}$) identified in  
\cite{tsuchiya+gouda+konishi_1996}.
While it is possible that there are different
time scales associated to different physical
processes as argued in
\cite{tsuchiya+gouda+konishi_1996}, it seems a 
more plausible interpretation to us to suppose that there is single
physical relaxation process leading,
albeit very slowly, to macroscopic relaxation 
of the system, and to characterize this
relaxation by a function and the scaling
of its parameters with $N$.
In this respect it is interesting to
note that  the specific stretched
exponential form we fitted to the temporal behaviour
has the known property \cite{montroll+bendler_1984}
that it is can be written as a weighted integral over 
simple exponentials (i.e. it can be interpreted
as arising from the superposition of an 
infinite number of relaxation processes
each with a single characteristic time).

In \cite{tsuchiya+gouda+konishi_1997} Tsuchiya et al. have also 
described chaotic ``itinerant" behaviour of 
these systems, starting from the same ($R_0=1$) initial 
conditions i.e., in which the system shows apparently stochastic
{\it macroscopic} behaviour. In our analysis  this would
correspond to such behaviour for the parameters $\phi_{11}$ 
or $\phi_{22}$. While we have seen that there are indeed
very significant fluctuations in these parameters, which
correspond to very significant differences in the ``macroscopic"
evolution of these systems, we have studied carefully
their dependence on $N$ and found them to decay
monotonically. The results of \cite{tsuchiya+gouda+konishi_1997}
were obtained for $N=64$, a range in which we still 
see fluctuations of $\phi_{11}$ which are order unity.
Only when we reach $N$ of order several hundred do
we see these fluctuations diminish significantly so
that the macroscopic trajectory of the system becomes
quite localized. We thus believe that as $N$ increases
these effects will becomes negligible, even on the 
time scales on which relaxation occurs, and an
effectively deterministic macroscopic evolution
will occur. 

It is interesting to compare our results also to those
of Yawn and Miller \cite{yawn+miller_2003}, who have analyzed 
in detail relaxation in a version of the sheet model in which 
there are sheets of different masses. In this case the relaxation
towards thermal equilbrium may be clearly distinguished 
by testing for equipartition of the kinetic energy, and the 
associated spatial segregation of the mass populations.
In simulations starting from waterbag type initial conditions 
with a virial ratio of two, for a range of different mass ratios 
and up to $N=128$ particles, clear evidence was found 
in \cite{yawn+miller_2003} for such relaxation using 
appropriately defined indicators. Like the order parameters
we have employed here, these show characteristic behaviours
corresponding to the principal phases of the dynamical
evolution (violent relaxation, QSS phase, relaxation towards
thermal equilibrium). Although we cannot 
compare our results directly, we note that the time scales
observed for relaxation of systems with $N \sim 10^2$
particles are quite consistent with those we have
observed for the equal mass system with initial virial
ratio $R_0=1$.  Yawn and Miller \cite{yawn+miller_2003}
also measure temporal  correlation properties and find
weak but persisting correlations characterized by a 
power-law decay (in time), which they interpret as evidence
for the incompleteness of relaxation. In the present study
we have found, in contrast, that our principal observables decay in 
time with a functional form which allows the identification
of characteristic time scales. Further all deviations of these 
observables from their equilibrium values decrease clearly 
as $N$ increases, and thus we have interpreted  the associated 
``incompleteness''  of relaxation simply in terms of finite 
$N$ effects. It would be interesting certainly to perform a
more direct comparison of the results in the two models,
and in particular to extend the study of Yawn and Miller
to allow a determination of the $N$ dependence of 
the parameters they study. We note also that Yawn and Miller
argue that the power-law decay suggests the existence of a 
fractal structure in phase-space, which, as we have been mentioned above, is also
proposed as an explanation in \cite{almeida_etal_2001} for
the appearance of relaxation characterized by a stretched
exponential behaviour.  

Let us finally mention some other issues of importance
concerning aspects of the dynamics of this system which
have been treated elsewhere but which we have not
discussed here. As we have discussed, we interpret
our results in line with those of many previous studies 
of  this and other long-range systems: the evolution from
an arbitrary out of equilibrium initial condition is characterized
a first phase of relaxation to a QSS, interpreted as a finite
particle sampling of a stationary solution of the Vlasov
equation, on a time scale independent of $N$, followed
by a slow relaxation to thermal equilibrium on an
$N$-dependent time scale.  Studies of the single mass 
sheet model for other specific initial conditions suggest
that this simple scheme may be too limiting, for this
model (and possibly, for all such models). On the
one hand Reidl and Miller have reported numerical
results \cite{reidl+miller_1995} for specific ``two cluster"
initial conditions which show a dependence on $N$
in the time scale for relaxation to a QSS. On the 
other hand,  as mentioned in the introduction, Rouet et al.
\cite{rouet+feix, mineau+rouet+feix_1990} have 
shown, using both particle simulations and simulations
of the Vlasov equation, for yet other initial conditions that 
``holes" which rotate in phase space may be present 
after violent relaxation and persist on very long time
scales. Although it is not  evident that there is necessarily 
a relation between either finding and the mechanism of relaxation 
to thermal equilibrium, a study incorporating 
such initial conditions would certainly be more complete that
that reported here. Extension of the study reported here to
larger $N$ still would likewise be desirable, despite the
extremely rapidly growing numerical cost of such 
simulations with $N$. 


\subsection{Acknowledgements}

The simulations were carried out in large part at the Centre de Calcul
of the Institut de Physique Nucl\'eiare et Physique des Particules.
We are particularly grateful to Laurent Le Guillou for advice and help.
We thank also Duccio Fanelli for providing us with his own code which
allowed us to perform checks of ours.  We thank P. Astier, J. Barr\'e, A. Gabrielli,
B. Marcos, P. Viot, F. Sicard, F. Sylos Labini for useful conversations,
comments or suggestions.  

\vskip 1cm

\bibliographystyle{aipproc}







\end{document}